\documentclass[twocolumn, showpacs, preprintnumbers, amsmath, amssymb, prapplied, reprint,nofootinbib,showkeys]{revtex4-1}
\usepackage{graphicx}
\usepackage{color}
\usepackage{amssymb}
\usepackage{amsmath}
\usepackage{mathrsfs}
\usepackage[capitalize]{cleveref}
\usepackage{multirow}

\begin{document}

\title{Optomechanical Gigahertz Oscillator made of a Two Photon Absorption free piezoelectric III/V semiconductor}

\author{In\`es Ghorbel$^{1,2}$}
\email{Corresponding author: ines.ghorbel@thalesgroup.com}
\author{Fran\c cois Swiadek$^{1}$, Rui Zhu$^{2}$, Daniel Dolfi$^{1}$, Ga\"elle Lehoucq$^{1}$, Aude Martin$^{1}$}
\author{Gr\'{e}gory Moille$^{1}$}
\altaffiliation[current affiliations: 1- Physical Measurement Laboratory, National Institute of Standards and Technology, Gaithersburg, MD, USA, 2- Maryland NanoCenter, University of Maryland, College Park, MD, USA ]{}
\author{Lo\"ic Morvan$^{1}$, R\'emy Braive$^{2,3}$, Sylvain Combri\'e$^{1}$, Alfredo De Rossi$^{1}$}
\affiliation{ $^{1}$Thales Research and Technology, Palaiseau, France}
\affiliation{ $^{2}$Centre de Nanosciences et de Nanotechnologies, CNRS, Universit\'e Paris-Sud, Universit\'e Paris-Saclay, C2N, Palaiseau, 91767, France}
\affiliation{$^{3}$Universit\'{e} Paris Diderot, Sorbonne Paris Cit\'{e}, 75207 Paris Cedex 13}

\date{\today}

\begin{abstract}
Oscillators in the GHz frequency range are key building blocks for telecommunication and positioning applications. Operating directly in the GHz while keeping high frequency stability and compactness, is still an up-to-date challenge. Recently, optomechanical crystals have demonstrated GHz frequency modes, thus gathering prerequisite features for using them as oscillators. Here we report on the demonstration, in ambient atmospheric conditions, of an optomechanical oscillator designed with an original concept based on bichromaticity. This oscillator is made of InGaP, a low loss and TPA-free piezoelectric material which makes it valuable for optomechanics. Self-sustained oscillations directly at 3 GHz are routinely achieved with a low optical power threshold of 40 $\mu W$ and short-term linewidth narrowed down to 100 Hz in agreement with phase noise measurements (-110 dBc/Hz at 1 MHz from the carrier) for free running optomechanical oscillators. 
\end{abstract}

\maketitle
\section*{Introduction}
Optomechanical (OM) resonators, exploiting the interaction between light and a moving optical cavity \cite{Kippenberg2008}, have been actively looked into in recent years with impressive demonstrations in the quantum regime \cite{Riedinger_2018,Cohen_2015,Purdy_2017}. Meanwhile, other important applications have also been found for ultra-compact sensors \cite{Krause2012_accelero}, microwave  to  optics  transduction \cite{Balram2016}, radiofrequency signals amplification \cite{massel_2011} or stable microwave oscillators \cite{Hossein-Zadeh2010}. Essential feature in modern navigation, communication and timing systems, microwave oscillators at high frequencies are compared in the light of their stability at their natural frequency and their form-factor. 
With their micrometric size and their mechanical resonance frequency already in the GHz range, OM crystals \cite{Eichenfield2009} (OMC) present a unique potential to reach ultra-compact stable microwave oscillators. OM oscillators have been investigated but still lie far from the microwave domain and spectral purity is, for the moment, an issue which has been scarcely addressed.\\
Besides, a shared limitation for every application is thermo-optical instabilities which limit the optical power injected inside the resonator. First OM resonators, and especially OMCs, made of silicon, suffer from two-photon absorption preventing quantum regime in cooling experiments to be achieved. Hence, different materials such as Silica \cite{Enzian_2019}, Silicon Nitride \cite{davanco_si3n4_2014} and diamond \cite{Burek2016,OM_single_diamond_crystal} have been considered as materials of choice thanks to their large thermal conductivity and low optical absorption. Thus, a high number of intracavity photons has been reached with diamond OMC \cite{Burek2016}. None of these materials shows piezoelectric properties which could efficiently bridge microwave to optics. Thus, they are unsuitable for hybrid opto-electro-mechanical devices \cite{Midolo_2018}, particularly attractive in various contexts, from telecommunications to quantum information and from classical radar to quantum radar \cite{Barzanjeh2015}. That is why non centro-symmetric crystals such as large electronic bandgap III-V semiconductors are appealing for optomechanics and have been recently investigated (Gallium Phosphide \cite{Schneider2018,GaP_microdisk} and Aluminium Nitride \cite{Bochmann2013}) as they do not suffer from Two Photon Absorption (TPA) when operating in the practical telecom spectral range.\\
Here we consider another material, Indium Gallium Phosphide (In$_{0.5}$Ga$_{0.5}$P) grown on GaAs. Owing to a large electronic forbidden gap ($\approx 1.9eV$) two-photon absorption is suppressed at telecom wavelengths \cite{combrie2009high}, which allows reaching a very large optical energy density and triggers nonlinear effects such as soliton pulse compression \cite{colman2010temporal}. For these reasons, InGaP has been introduced recently in optomechanics \cite{Buckle_2018,Cole2014_GaInP,Guha2017}, but an OMC have not been realized yet. We introduce a new design concept, relying on bichromaticity \cite{combrie2017compact}, which presents the advantage of being robust to fabrication disorder \cite{delphin_2018} and thus achieved systematically functional devices with large optical Q factors and low mechanical losses. The self-sustained oscillation has been characterized in detail all the way to the measurement of the phase noise, revealing that our OMC is comparable to much larger microtoroids made of Silicon Nitride. 

\section*{Cavity design and modeling}
\begin{figure}
\includegraphics[width=\columnwidth]{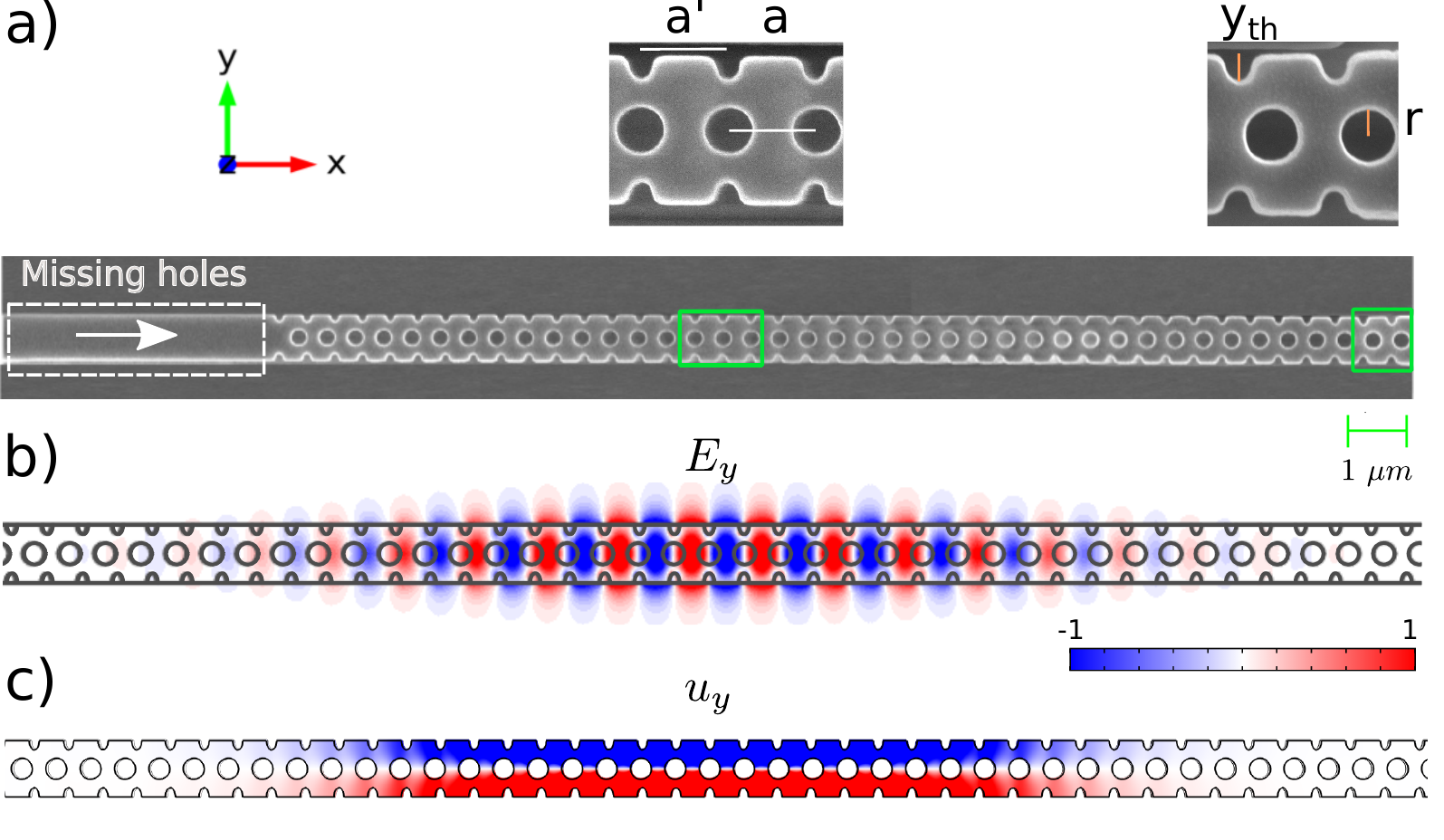}
\caption{\label{fig:Fig1} a) Scanning Electron Microscope image of the fabricated structure; the area where holes are removed is delimited by a dashed white line; white arrow: input light b) calculated normalized real part of the field $E_y$ for the fundamental optical mode at $\lambda$=1545 nm ; c) calculated normalized mechanical displacement $u_y$ at $f_{m}=3.12 GHz$.} 
\end{figure}

The widespread designs introduced in \cite{Safavi-Naeini2011,gomis2014one} rely on tapering the crystal parameters, following a well optimized profile, according to the concept of ``gentle confinement'' \cite{Song_2005}. Our design is based on a radically different concept, which does not use any tapering at all: all holes are the same with constant radius $r$ and period $a$ while the sidewall modulation has a constant depth $y_{th}=0.27a$ and is strictly periodic with period $a^\prime$. The two periods are however slightly different, $a^\prime=0.98 a$. As shown by \cite{alpeggiani_effective_2015} in the context of two dimensional photonic crystals, this creates an effective confining potential which minimizes radiative leakage but still keeps the mode volume low. Thus, the design is described by only 4 parameters and requires no optimization as the radiative leakage limited Q of the fundamental mode, calculated by Finite Difference Time Domain method, is always above $10^6$ as $r$ and $y_{th}$ are varied over a fairly broad range (see Supplementary Information), which also suggests robustness against fabrication tolerances. The next optical mode is located about 2 THz below in the spectrum (see supplementary).\\
The implementation of this concept in the context of optomechanics also requires that the same structure also confines mechanical modes. To the best of our knowledge, the possibility of localizing a mechanical mode using a bichromatic structure has not been considered. The mechanical modes in Fig.~\ref{fig:Fig1}c are computed using the Finite Element Method, implemented in the COMSOL software. The confinement of the mechanical breathing mode oscillating at about 3 GHz (Fig.\ref{fig:Fig1}c) is explained by the local increase of the stiffness in the structure induced by the increasing misalignment of holes and sidewalls as moving outwards from the center of the cavity. The fundamental mode has the highest frequency (see supplementary information for mechanical spectrum). \\
The calculated optical mode volume $V_{opt}$, the effective oscillator mass $m_{eff}$, the mechanical mode volume $V_m$ and the vacuum optomechanical coupling constant $g_0$ depend on the parameter $a^\prime/a\approx1$, providing a simple ``knob'' for tuning the device properties (details in supplementary). The largest $g_0$ involves the fundamental optical and mechanical modes, any other combination of modes results in a much smaller coupling. The calculated photoelastic\footnote{as data for InGaP is absent in the literature we have used the parameters of GaP as in \cite{Guha2017}} and moving boundary contributions\cite{Baker_2014} are: $g_{0,MB}/2\pi=-117kHz$ and $g_{0,PE}/2\pi=494kHz$, hence  $g_{0}/2\pi=377 kHz$ (see supplementary for details on $g_0$ computation and the values of the photoelastic tensor).

This design ensures the simultaneous localization of photons (Fig.~\ref{fig:Fig1}b, $V_{opt}=0.97(\lambda/n)^3$) and phonons (Fig.~\ref{fig:Fig1}c, $V_m=2.5\times 10^{- 19} m^{3}$, $m_{eff}=1.01 pg$). The cavity is coupled to the input waveguide by removing $n_{lh}$ holes and sidewall corrugation on one side (out of the 51 holes in total) and the waveguide is coupled to a lensed fiber using an inverse taper~\cite{tran2009photonic}. \\ 

\begin{figure}
\includegraphics[width=\columnwidth]{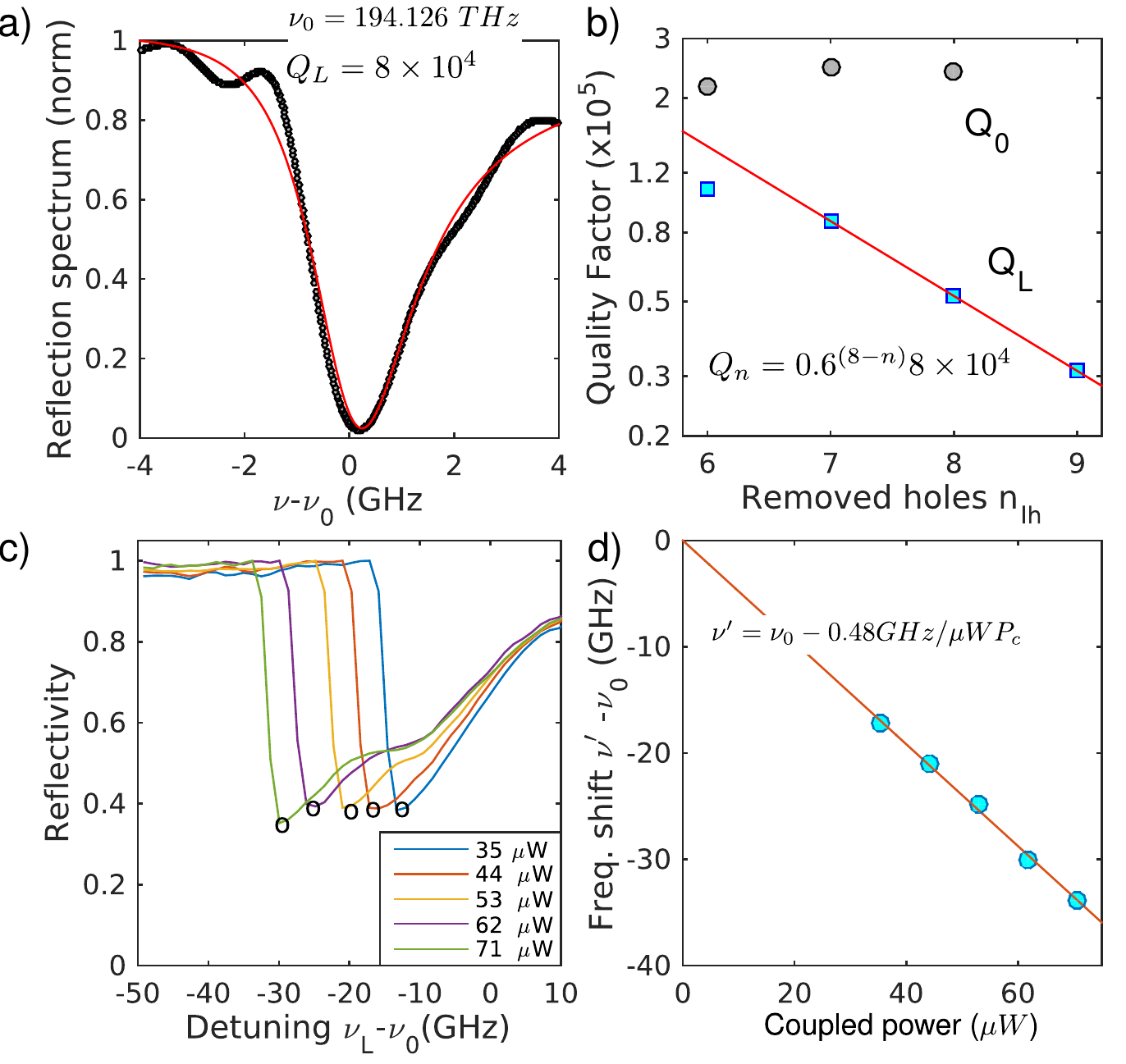}

\caption{\label{fig:Fig2} Fundamental mode: a) Reflection spectrum (black), Lorentzian fit (red) and extracted loaded Q factor for a cavity with 7 holes removed; b) measured loaded and intrinsic Q-factors as a function of the number of holes removed, and exponential fit (red line); c) normalized reflectivity as a function of the detuning; d) extracted thermo-optic shift as a function of the on-chip power $P_c$ for a cavity with 9 holes removed.}
\end{figure}

The device is fabricated on an InGaP membrane grown by MOCVD lattice-matched to GaAs. The OM crystal is processed following the same recipe as for two dimensional photonic crystals\cite{combrie2017compact,combrie2009high}.
\section*{Optical characterization}
The optical resonances are probed in a reflection geometry using a high resolution optical heterodyne technique\cite{combrie2017compact}. This provides access to the complex spectrum of the cavity (see supplementary). Its modulus is shown in Fig.~\ref{fig:Fig2}a. We consider the mode in the spectrum with the highest frequency (the fundamental). The loaded quality factor $Q_L$ decreases by a factor 0.6 for each period removed, while the intrinsic quality factor $Q_0$, extracted from the fit of the measured complex amplitude, is $2.2\pm0.2\times10^5$ (Fig.~\ref{fig:Fig2}b)\footnote{the procedure used here does not operate in extreme conditions such as overcoupled cavity. }.We measured an intrinsic quality factor over $10^5$ in 9 out of 12 nominally identical cavities. The whole spectrum is shown in Fig.9 of the supplementary information where the first order mode can be seen. \\
Absorption, at room temperature, is extracted from the normalized reflectivity as a function of the laser detuning $\nu_L-\nu_0$ swept from blue to red such that the resonance is thermally pulled \cite{Carmon2004} until the bistable transition occurs (Fig.~\ref{fig:Fig2}c). This, to a very good approximation, corresponds to the detuned resonance $\nu^\prime$ (see supplementary information). When plotted against the on-chip power (i.e. the incident power), $\nu^\prime$ reveals a linear dependence (Fig.~\ref{fig:Fig2}d), hence suggesting linear absorption, likely due to defects at the surface. Following the same procedure as in \cite{martin2017gainp},  the dissipated power is extracted based on the calculated thermal resistance and the measured dependence of the resonance with temperature. This leads to an estimate of the absorption rate $\Gamma_{abs}/2\pi=8\,MHz$, which is much smaller than the total intrinsic losses $\Gamma_0/2\pi\approx 1 GHz$. Correspondingly, the fraction of the dissipated on-chip power is $\alpha=4\Gamma_{abs}(\kappa-\Gamma_0)/\kappa^2\approx0.4\%$, with $\kappa$ the photon cavity decay rate. Absorption could be interpreted in terms of an effective imaginary refractive index\footnote{which should depend on the geometry since it represents absorption due to surface defects.} through $n^{\prime}(InGaP)=n(InGaP)\Gamma_{abs}/2\pi\nu\approx10^{-7}$, which is substantially lower than the estimate in \cite{Cole2014_GaInP} at $\lambda=$1064nm and consistent with measurement of intrinsic $Q>10^6$ still limited by elastic scattering\cite{combrie2017compact}.\\

\begin{figure}
\includegraphics[width=\columnwidth]{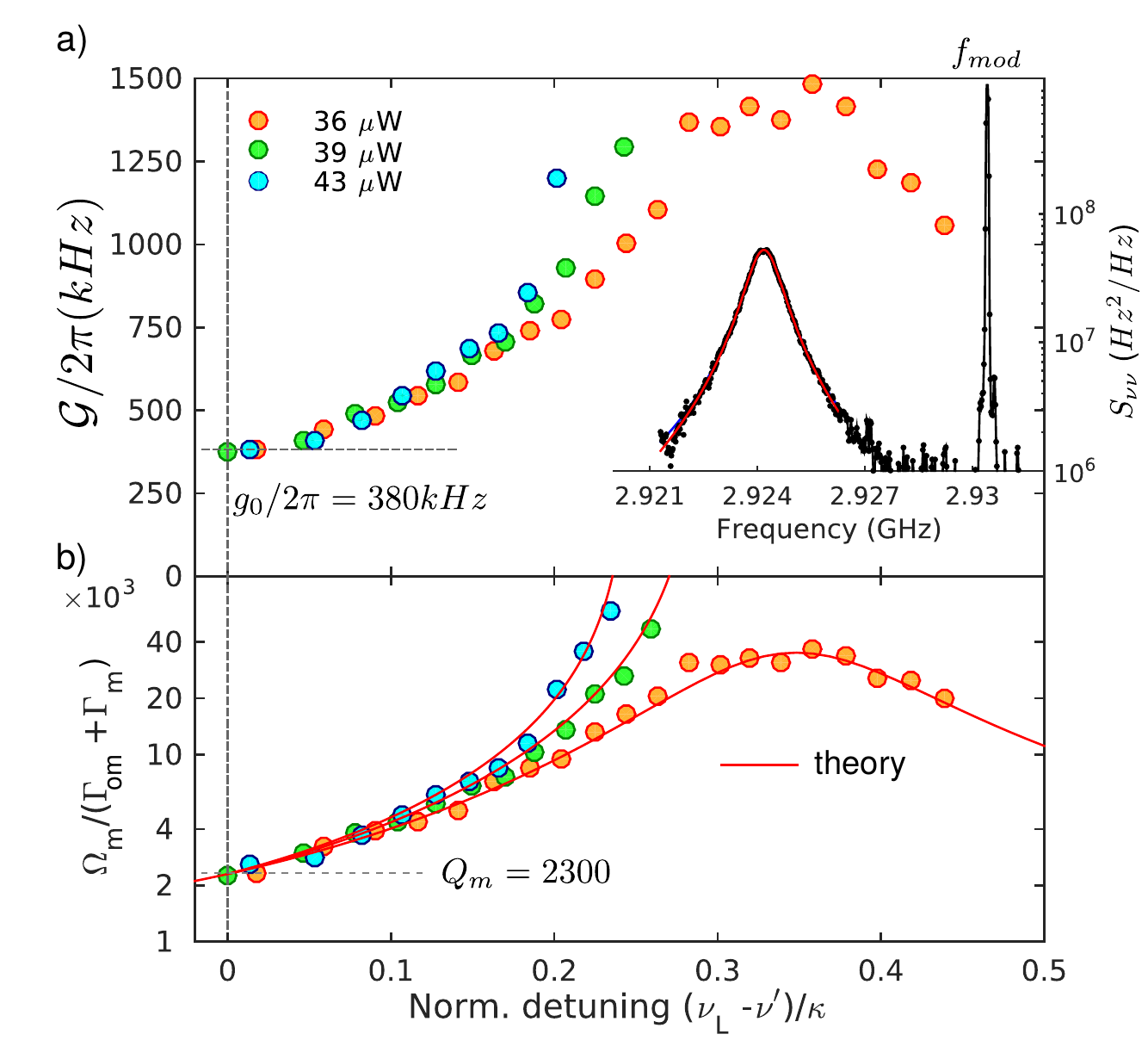}
\caption{\label{fig:Fig3} a) Measured vacuum optomechanical coupling as a function of the normalized laser detuning for 3 different on-chip laser power; inset : calibrated power spectral density of the frequency fluctuation along with calibration tone $f_{mod}$ and Lorentzian fit. b) corresponding measured mechanical linewidth compared to theory.}
\end{figure}
                                       
\section*{Probing of Brownian motion of the oscillator}
The optomechanical crystal considered in this section* has an optical quality factor of $Q=3 \times10^4$. The noise spectrum of the mechanical resonator reveals several peaks. The one with the largest frequency ($f_m=2.924\ GHz$, see inset Fig\ref{fig:Fig3}a) is identified as the fundamental mode (see Fig.10 of the supplementary for mechanical spectrum). The vacuum optomechanical coupling is measured at room temperature and standard pressure with the technique discussed in ~\cite{gorodetksy2010determination}. The reflected optical power is detected by a fast Avalanche Photodiode which is amplified by a 40dB low noise amplifier before going to an electric spectrum analyser (ESA). The electric power spectra corresponding to the mechanical motion of the resonator is compared to a calibration tone with spectrum $S_{mod}$ generated by a phase modulator in the input optical path, allowing the measurement of the power spectrum of the frequency modulation $S_{\nu\nu}$, as shown in the inset of Fig.~\ref{fig:Fig3}a (details in supplementary). In our case, it was not possible to operate the OM resonator at low enough power to avoid dynamical backaction while maintaining the detection level well above noise. Thus, what we measure and plot on Fig.~\ref{fig:Fig3}a is a quantity $\mathcal{G}=\sqrt{\frac{\int{S_{\nu\nu}(f)df}}{n_{th}}}$ that corresponds to $g_0$ at vanishing  laser-cavity detuning $\nu_L -\nu^\prime$, which is corrected for the thermally induced spectral shift, see supplementary. 
Considering the uncertainty on the photoelastic coefficients, the measured $g_0/2\pi=380$ kHz is very close to the calculations solely including the photoelastic and moving boundary contributions. This is consistent with the fact that the thermo-mechanical term\cite{Guha2017} is negligible in our system (discussion in Supplementary).\\
The corresponding mechanical linewidth (Fig.~\ref{fig:Fig3}b) is measured and compared to theory \cite{Aspelmeyer2014} accounting for the narrowing due to the dynamical backaction $\Gamma_{om}$, when $\Delta=\nu_L -\nu^\prime>0$:
\begin{align*}
\Gamma_{om}&=n_{h\nu}g_{0}^{2} \times\\
 &\times\left[\frac{\kappa}{(\Delta+ 2\pi f_{m})^{2}+\kappa^{2}/4}
 -\frac{\kappa}{\left(\Delta - 2\pi f_{m}\right)^{2}+\kappa^{2}/4} \right]
\end{align*}
with the number of photons in the cavity given by the usual coupled mode theory.

The parameters used in the model (gathered in a table in the supplementary) are measured: $\kappa/2\pi=6.5$ GHz, $\Gamma_0/2\pi=0.9$ GHz, $\Omega_m/2\pi=2.92$ GHz and $g_0/2\pi = 380$ kHz. Only the on-chip laser power levels used in the model, $P_c= 43.5$, $47.9$ and $51$ $\mu W$, have been adjusted within 20\% of the experimental values indicated in Fig \ref{fig:Fig3}a. From the lorentzian fit in the inset of Fig \ref{fig:Fig3}a, the mechanical linewidth is equal to $\Gamma_m/2\pi=1.2 MHz$ and the mechanical Q factor at room temperature and atmospheric pressure is $Q_m=\Omega_m/\Gamma_m=2300 \pm 150$ corresponds to the measurement at zero detuning.\\

\section*{Self-sustained oscillations}

\begin{figure}
\includegraphics[width=\columnwidth]{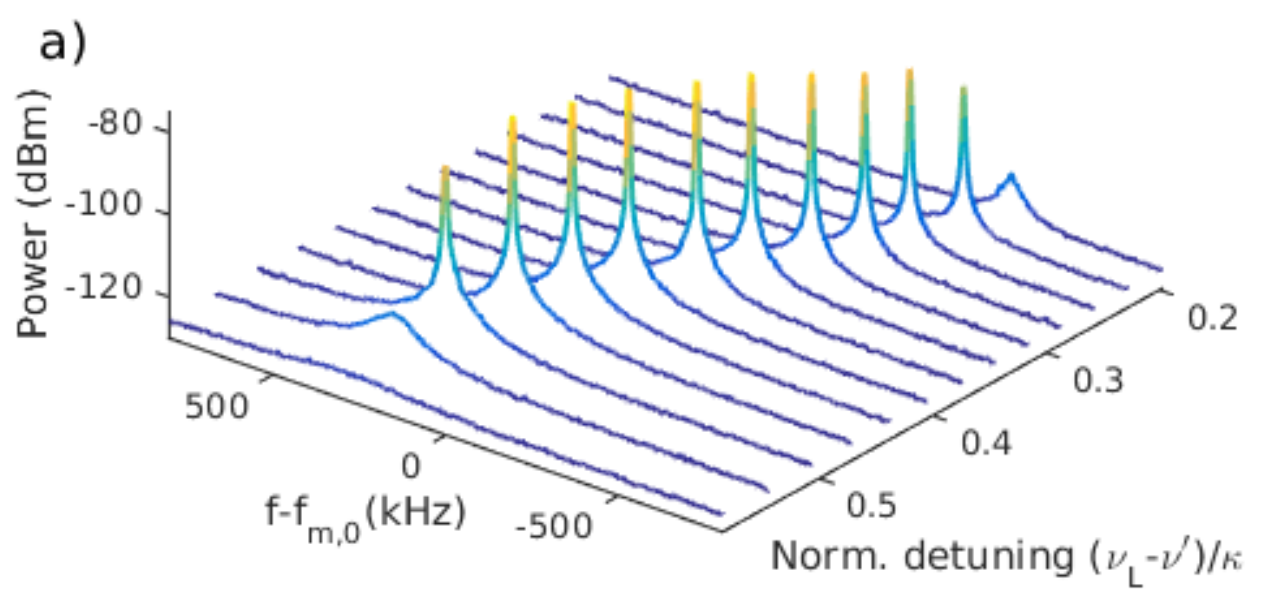}
\includegraphics[width=\columnwidth]{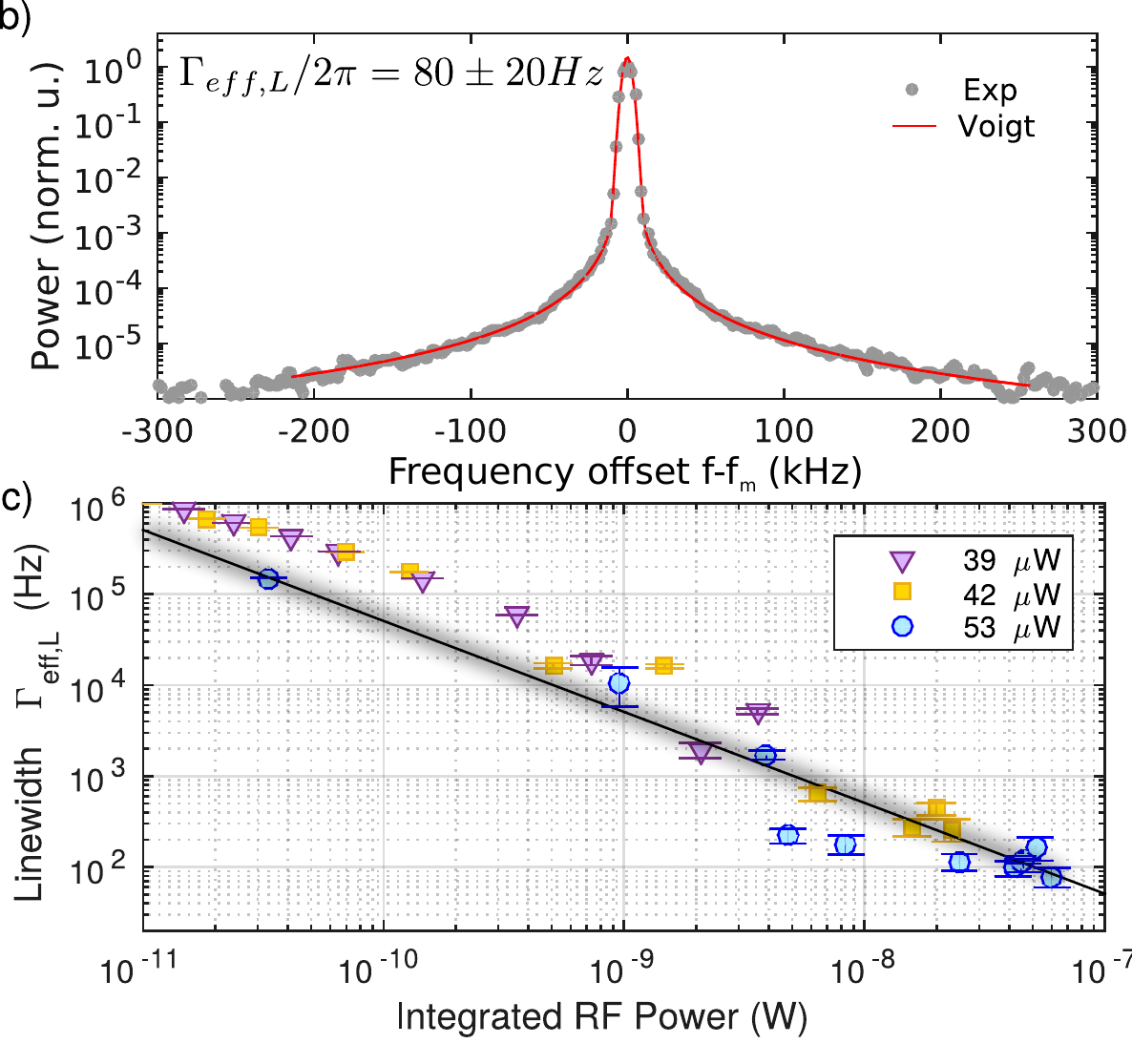}
\caption{\label{fig:Fig4} a) Raw spectra of the detected signal as a function of the detuning for $P_c=53\mu W$; b) Fit of normalized RF spectrum with the Voigt function; c) fitted Lorentzian linewidth $\Gamma{eff,L}$ as a function of the RF integrated power for different optical pump levels, the black line represents the estimated short term limit based on eq.~\ref{eq:STth}, blur represents uncertainty on the measurement of the RF power at thermal equilibrium.}
\end{figure}

We routinely observe self-sustained oscillations on devices with different loaded Q factor. We focus on the cavity with loaded quality factor $Q=30\ 000$. As the power is increased, the resonator eventually undergoes regenerative oscillations. The threshold is predicted by the condition that the mechanical loss equates the optical anti-damping calculated above: $\Gamma_m+\Gamma_{om}=0$. Using the measured parameters above yields $P_{c,tr}=47\mu W$, which is again, within 20\% of the measured value, 40 $\mu W$.\\
The measurement is performed as the laser is swept towards the red across the resonance and repeated as the on-chip power is increased. Through the dynamical backaction, the mechanical mode drifts by 700 kHz for an on-chip power of $53 \mu W$ (Fig.~\ref{fig:Fig4}a). The mechanical linewidth is very well fitted by a Voigt function (Fig.~\ref{fig:Fig4}b) which is the convolution of a gaussian function which Full Width at Half Maximum is equal to $\sigma_{G} = 5047 \pm 929 Hz$ (which corresponds to the Resolution Bandwidth used to record the different spectra) and a lorentzian function. The lorentzian linewidth, corresponding to the short-term linewidth, decreases from $1.2 \pm 0.08$ MHz to $\Gamma_{eff,L}/2\pi=80 \pm 20 Hz$ for an on-chip power of $53 \mu W$. \\
On Fig.~\ref{fig:Fig4}c), the short-term linewidth is plotted against the RF integrated power. We consider the transduction of the mechanical movement to the optical signal to be constant and linear. In that case, the number of phonons $\overline{n}$ can be deduced from $\overline{n}_{th}/\overline{n}=P_{RF, th}/P_{RF}$ where $\overline{n}_{th}$ is the number of phonons at thermal equilibrium, given by $\overline{n}_{th}=k_BT/\hbar\Omega_m$ and $P_{RF,th}$ is the RF integrated power at thermal equlibrium, when there is no dynamical backaction. 
The knowledge of the number of phonons allows one to calculate the limit to the short-term linewidth given in \cite{Vahala2008backactionlimit,Hossein-Zadeh2010}, similarly to the Shawlow-Townes limit for lasers:\\
%
\begin{equation}
	\Gamma_{eff,L}=\Gamma_m \left( \frac{\overline{n}_{th}}{2\overline{n}}+\frac{1}{2\overline{n}} \right) \approx\Gamma_m\frac{\overline{n}_{th}}{2\overline{n}}
\label{eq:STth}
\end{equation}

Eq \ref{eq:STth} is valid above threshold and is plotted in black on Fig.~\ref{fig:Fig4}c). As the measurements are performed at room temperature, $\overline{n}_{th}+1 \approx \overline{n}_{th}$ and in that case, as pointed out in Ref.\cite{Hossein-Zadeh2010}, the short-term linewidth is limited by thermal noise. As the experimental points obtained by fitting the spectra with the Voigt function follow the limit given by eq.\ref{eq:STth}, we can conclude that the short-term linewidth of the self-sustained oscillations is limited by Brownian motion and this should be improved by lowering the temperature bath.

\begin{figure}
\includegraphics[width=\columnwidth]{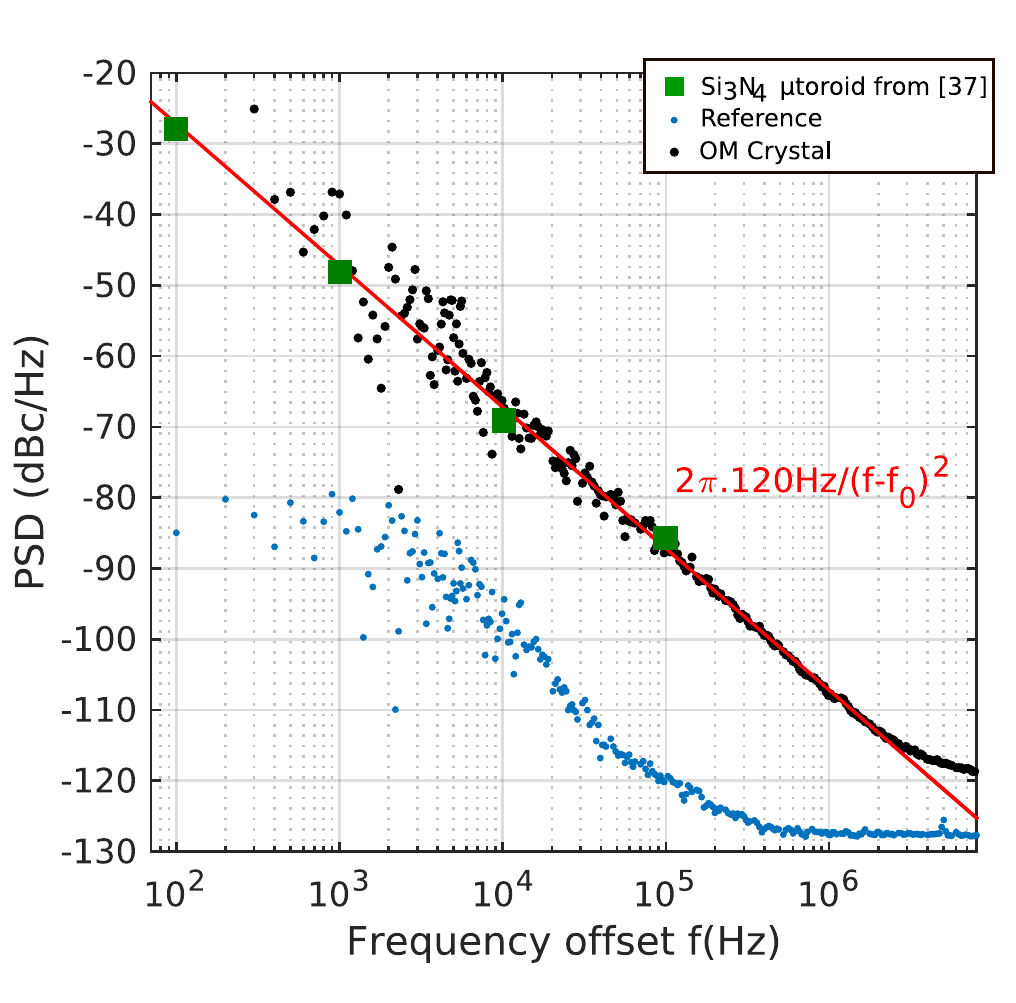}
\caption{\label{fig:Fig5}  Measured phase noise spectrum (black filled circles), reference (blue filled circle), phase random walk noise corresponding to a Lorentzian linewidth $\Gamma_{eff,L}/2\pi=120 Hz$ (red line) and corrected phase noise of the silicon nitride microtoroid from \cite{Tallur_2011} (green squares)}
\end{figure}

A deeper insight in the noise properties of the oscillator\cite{Hossein-Zadeh2010} is gained by examining the spectral density of the phase noise $\mathcal{L}(f)$ (Fig.~\ref{fig:Fig5}), measured when the device is oscillating at its maximum amplitude. The cavity considered for this measurement has slightly different parameters (in particular a lower optical quality factor $Q=2.5\times10^4$) and a stronger Signal to Noise Ratio is obtained through optical heterodyning (see supplementary). From 5kHz to 2 MHz the phase noise spectral power density follows the slope $PSD=\Gamma_{eff,L}/f^2$, which is associated to phase random walk. The Lorentzian linewidth $\Gamma_{eff,L}/2\pi=120 Hz$ is extracted, which is consistent with the direct measurement on the signal spectral power (Fig.~\ref{fig:Fig4}). While white phase noise, due to thermal noise in the photodetector, dominates at higher frequencies, technical noise ($1/f^3$) dominates below 5kHz, which is typical of a free running oscillator. 

\section*{Conclusion}
In conclusion, an optomechanical crystal based on InGaP, a III-V piezoelectric semiconductor, has been developed based on a novel design involving only 4 parameters and requiring no optimization. The typical intrinsic optical Q factor is about $2\times10^5$, whereas the loaded Q is controlled by removing holes. While nonlinear absorption is absent in the telecom spectral range, owing to the large electronic band-gap, the linear absorption is very small ($\Gamma_{abs}/2\pi=8MHz$), which, combined to a long thermal relaxation rate compared to the oscillation frequency, implies a negligible contribution of thermomechanical forces to damping $\Gamma_{om}$.  The measured vacuum coupling constant is $g_0/2\pi\approx 380$ kHz, in good agreement with modeling. At room temperature and standard pressure, the mechanical damping is $Q_m=2300$, with a corresponding figure of merit $Q\times f=6\times10^{12}$, which is of the same order of magnitude as \cite{Cole2014_GaInP}. Self-sustained oscillations are achieved routinely with a loaded optical $Q_L> 2.5\times 10^4$, with an on-chip optical power level of about 40 $\mu W$. The measured mechanical short-term linewidth narrows down to about 100 Hz, limited by classical Brownian noise and would decrease with temperature. Compared to other optomechanical oscillators, the $1/f^2$ term of the phase noise is basically the same as in Silicon Nitride microtoroids\cite{Tallur_2011}, which is also a low loss material, once corrected for the carrier frequency to allow a fair comparison\footnote{$20log\left(N\right)+\mathcal{L}(f)$ where N is the ratio between the higher and lower operation frequency}. We note that Micro-Electro-Mechanical Systems (MEMS) \cite{Sridaran_2012} are about 10 dB below but our OMC provides an optical output, convenient for the distribution of the signal on-chip. Completed with piezo-electric transducers and hybridized on a Silicon Photonic circuit \cite{Integrated1652}, this device could be used for microwave to optical conversion and more elaborate miniaturized optoelectronic oscillators. We note that self-stabilisation schemes have been proposed for OM resonators\cite{Matsko2011_selfref}. Further improvement could be achieved by inducing tensile stress in the membrane \cite{Buckle_2018,Ghadimi_2018}. In perspective, this technology could be suitable for the investigation of complex non linear phenomena \cite{Sotomayor_non_linear_dynamics}, synchronization of several oscillators \cite{Heinrich2011} or quantum experiments.\\

\begin{acknowledgments}
This work was supported by the European Union's Horizon 2020 research and innovation programme under grant agreement No. 732894 (FET Proactive HOT).\\
This work was also partly supported by the RENATECH network
- We acknowledge support by a public grant overseen by the French National Research Agency (ANR) as part of the “Investissements d’Avenir” program: Labex GANEX (Grant No. ANR-11-LABX-0014) and Labex NanoSaclay (reference: ANR-10-LABX-0035) with Flagship CONDOR.
Authors declare no competing interests.
\end{acknowledgments}
\appendix

\section*{Calculated parametric dependence of radiation losses, volume and OM coupling}

\subsection{Dependence with hole radius}
\begin{figure}[ht!]
	\centering
	\includegraphics[width=0.5\columnwidth]{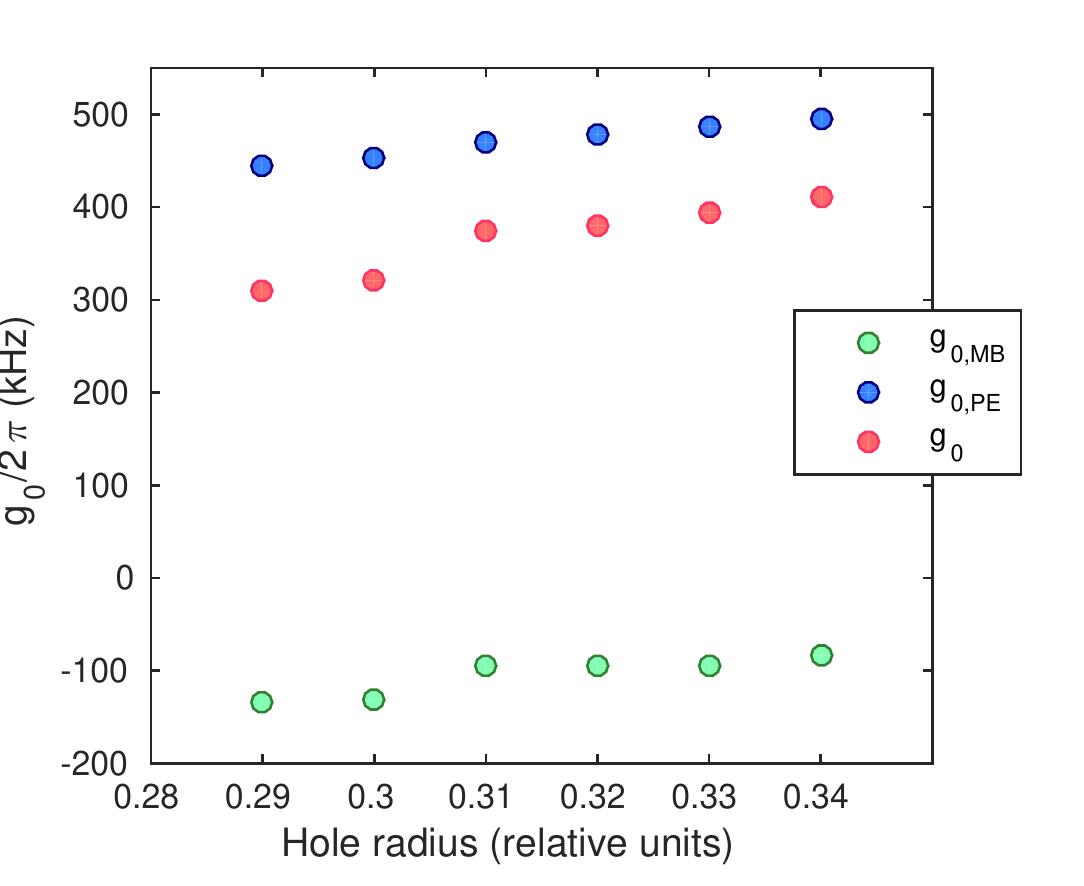}
	\caption{Evolution of the optomechanical coupling with the hole radius (units of a)}
	\label{fig:g0fctr}
	\centering
	\includegraphics[width=0.5\columnwidth]{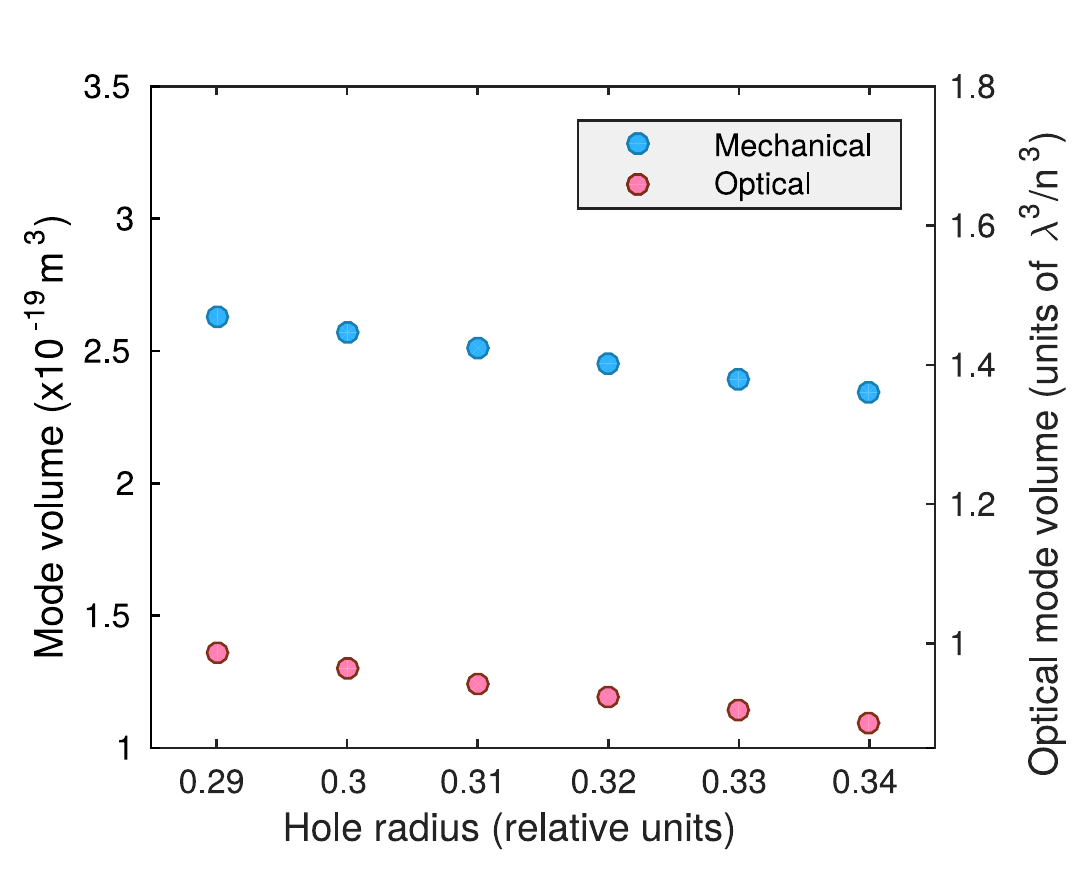}
	\caption{Evolution of the optical and mechanical volume with the hole radius (units of a)}
	\label{fig:Voptmfctr}
	\centering
	\includegraphics[width=0.5\columnwidth]{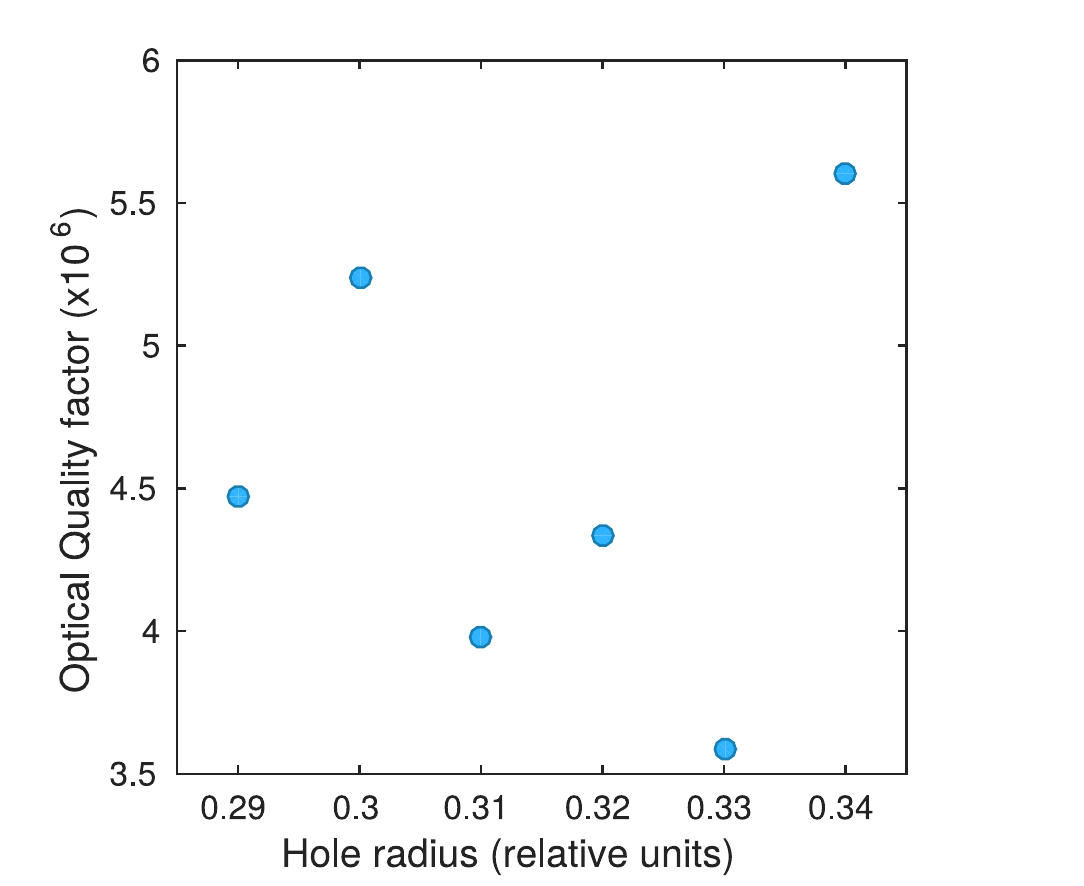}
	\caption{Evolution of the optical quality factor with the hole radius (units of a)}
	\label{fig:Qfctr}
\end{figure}
The radius of the hole is a diffcult parameter to control during fabrication. The graph in Fig \ref{fig:g0fctr} show a variation of as much as 100 kHz for the optomechanical coupling with holes of increasing values. 

\subsection{Dependence with teeth depth}
\begin{figure}[ht!]
	\centering
	\includegraphics[width=0.5\columnwidth]{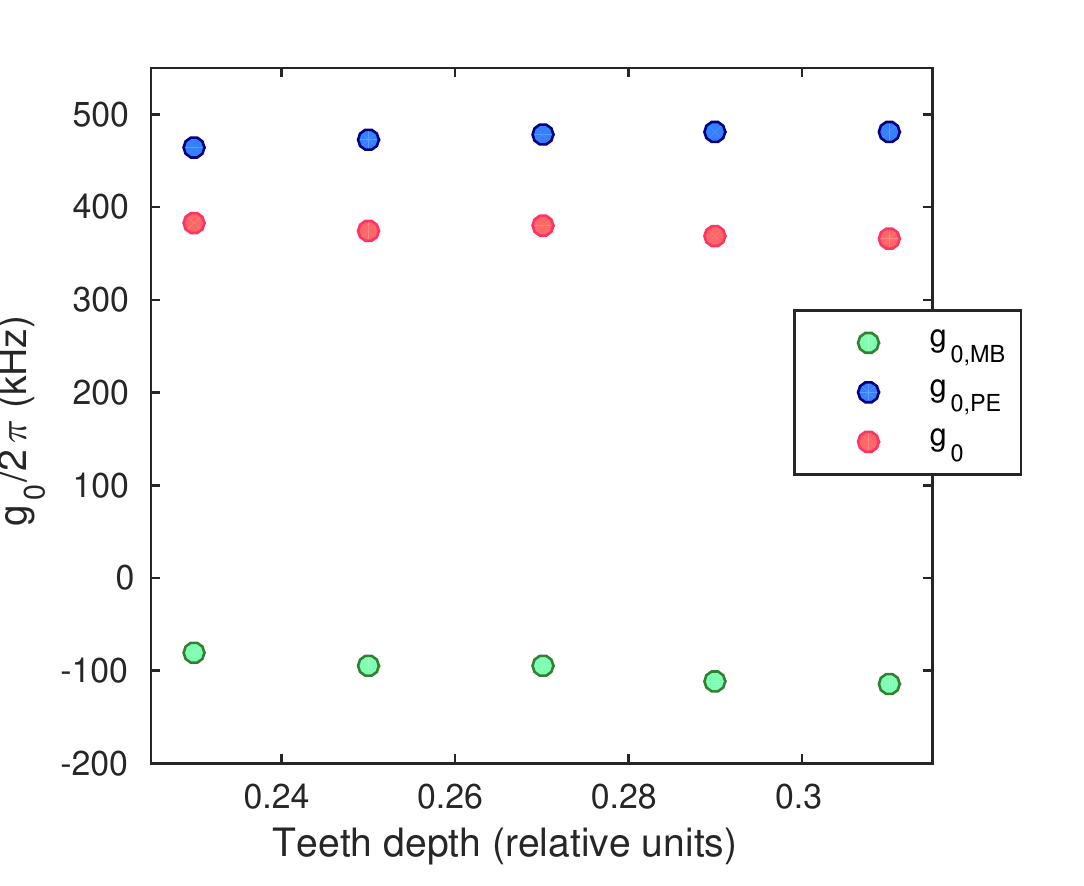}
	\caption{Evolution of the optomechanical coupling with teeth depth (units of a)}
	\label{fig:g0fctyth}
	\centering
	\includegraphics[width=0.5\columnwidth]{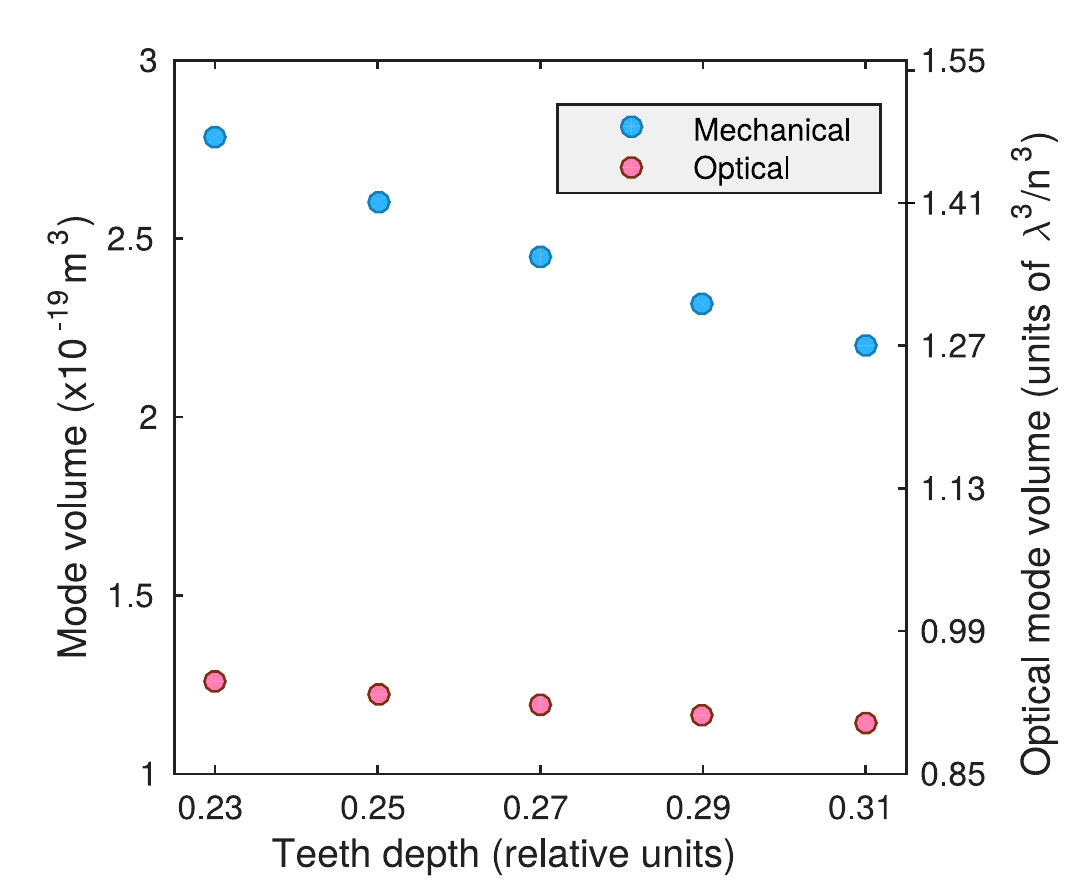}
	\caption{Evolution of the optical and mechanical volume with the teeth depth (units of a)}
	\label{fig:Voptmfctyth}
	\centering
	\includegraphics[width=0.5\columnwidth]{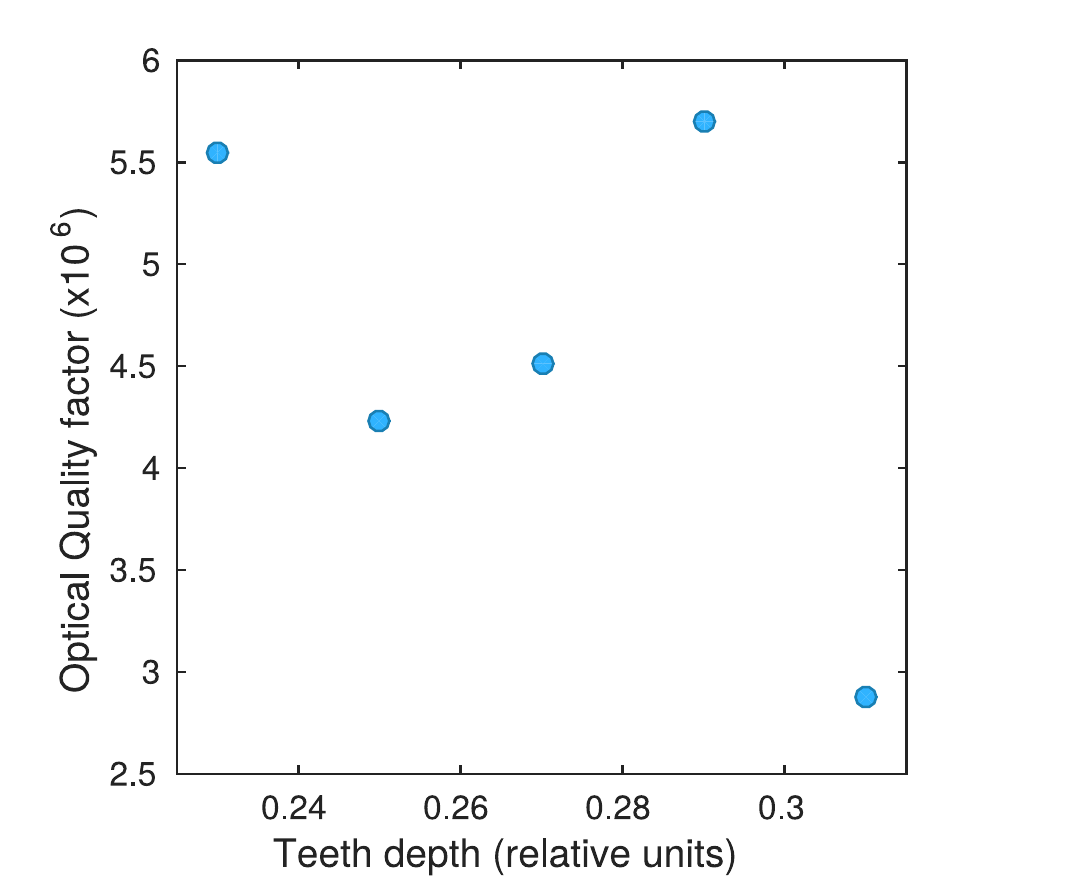}
	\caption{Evolution of the optical quality factor with the teeth depth (units of a)}
	\label{fig:Qfctyth}
\end{figure}
As can be seen on Fig\ref{fig:g0fctyth}, the depth of the teeth does not seem to have a significant influence on the optomechanical coupling $g_{0}$.\\
From figures \ref{fig:Qfctyth} and \ref{fig:Qfctr}, it can be seen that the optical quality factor does not depend on the hole radius and teeth depth when these paramters are changed over almost 30 nm, which suggests robustness against fabrication disorder.
\subsection*{Dependence with the ratio between the periods of the two lattices}
\begin{figure}[ht!]
	\centering
	\includegraphics[width=0.5\columnwidth]{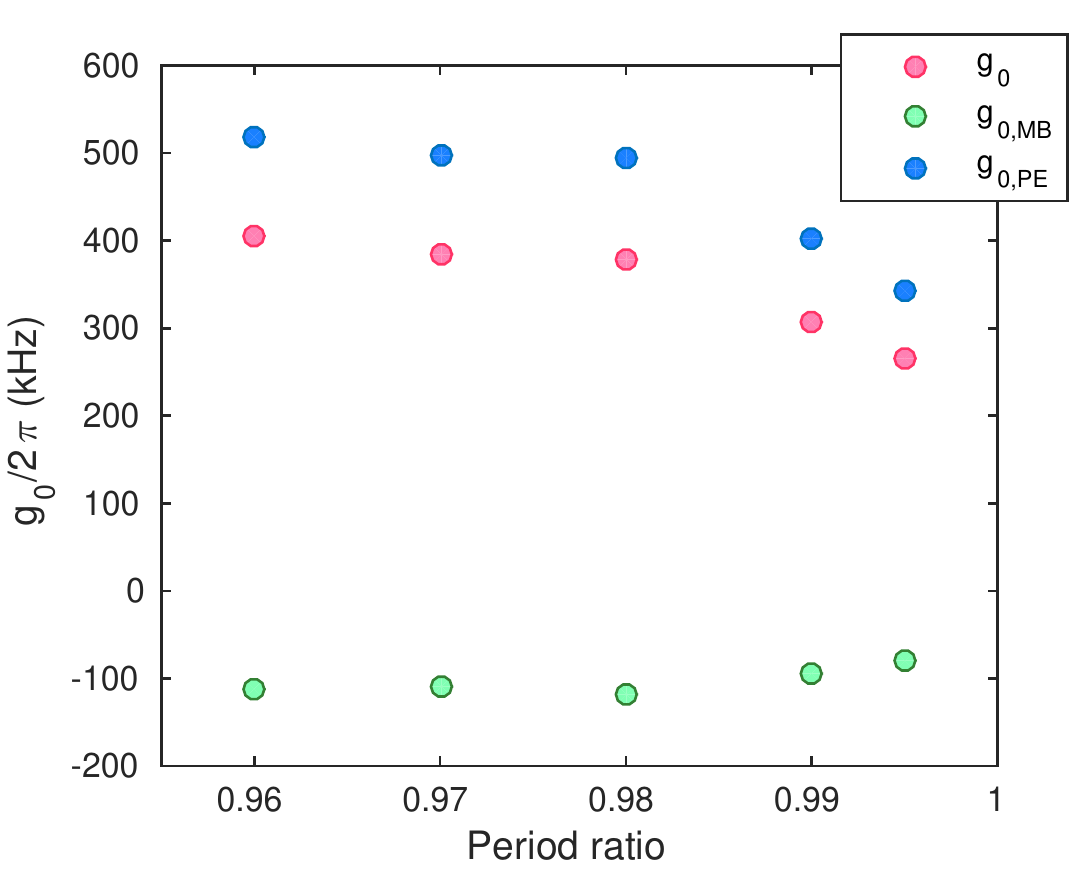}
	\caption{Evolution of the optomechanical coupling with the ratio of the periods (units of a)}
	\label{fig:g0fctar}
	\centering
	\includegraphics[width=0.5\columnwidth]{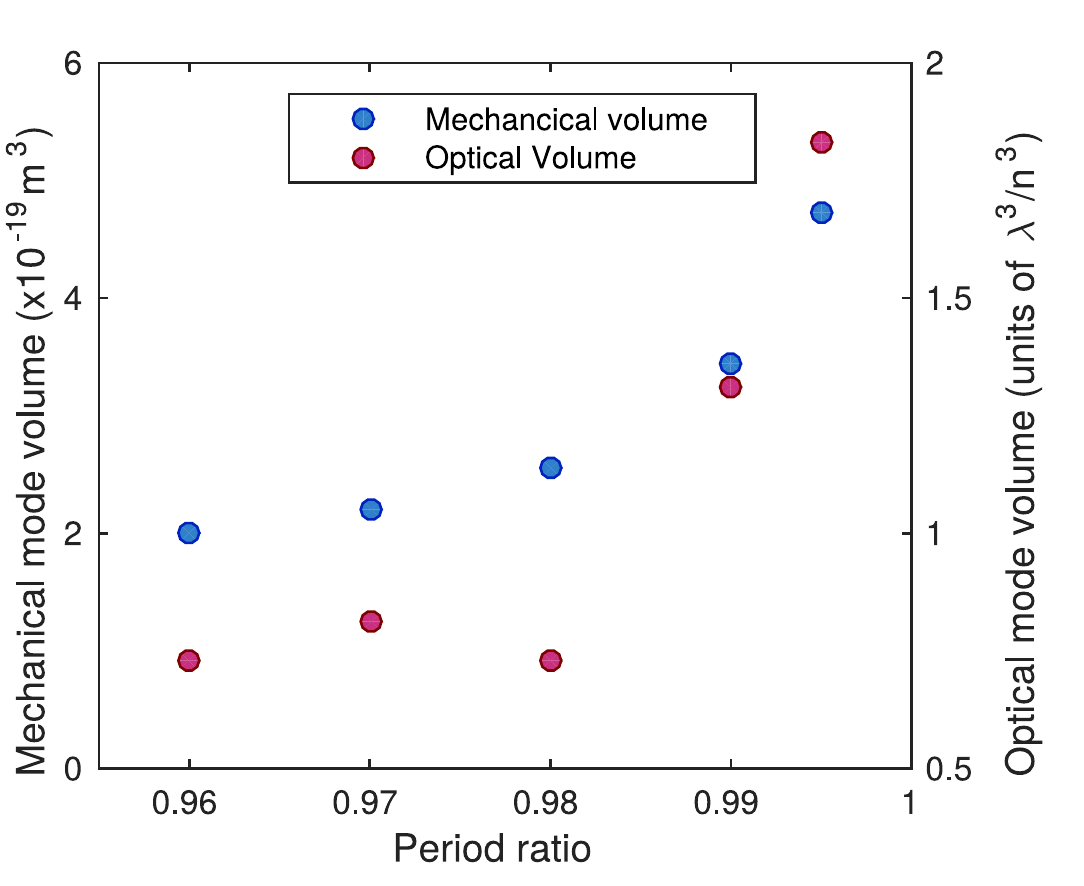}
	\caption{Evolution of the optomechanical coupling with the ratio of the periods (units of a)}
	\label{fig:Voptmfctar}
\end{figure}
From Fig.\ref{fig:g0fctar} and Fig.\ref{fig:Voptmfctar}, it is clear that as the ratio increases, the optical and the mechanical modes have a larger volume. Moreover, the optomechanical coupling $g_0$ is stronger when the mechanical and the optical modes are confined in a smaller volume. 
\appendix
\section*{Computing of vacuum OM coupling}
	The optomechanical coupling is calculated using the expressions from \cite{Balram2014} :
	\begin{eqnarray}
		g_{0,MB}&=-\frac{\omega_{0}}{2}\frac{\iint_S \overline{\overline{Q}}.\vec{n}\left(\Delta\varepsilon\left|\left|\vec{E}_{\parallel}\right|\right|^{2}-\Delta\varepsilon^{-1}\left|\left|\vec{D}_{\perp}\right|\right|^{2}\right)dS}{\iiint_V \varepsilon\left|\left|\vec{E}\right|\right|^{2}dV}\\
		g_{0,PE}&=\frac{\omega_{0}\varepsilon_{0}n^{4}}{2}\frac{\iiint_V\vec{E}\cdot p_{idkl}S_{kl} \cdot \vec{E}^{\ast}dV}{\iiint_V \varepsilon\left|\left|\vec{E}\right|\right|^{2}dV}
	\end{eqnarray}
\appendix
\section*{Photoelastic parameters}
	The photoelastic parameters for the computation of $g_{0}$ are taken from \cite{Mytsyk:15} :
	\begin{eqnarray*}
		p_{11}&=-0.23\\
		p_{12}&=-0.13\\
		p_{44}&=-0.10\\
	\end{eqnarray*}
	We note that there is an uncertainty of about $10\%$ on the above values. 
	These values are given for a null angle with the (001) axis. As the injection axis of our cavities is along (110), a rotation must be applied to the photoelastic tensor.\cite{Balram2014}.
\appendix
\section*{Optical spectrum}
\begin{figure}[ht!]
\centering
\includegraphics[width=0.9\columnwidth]{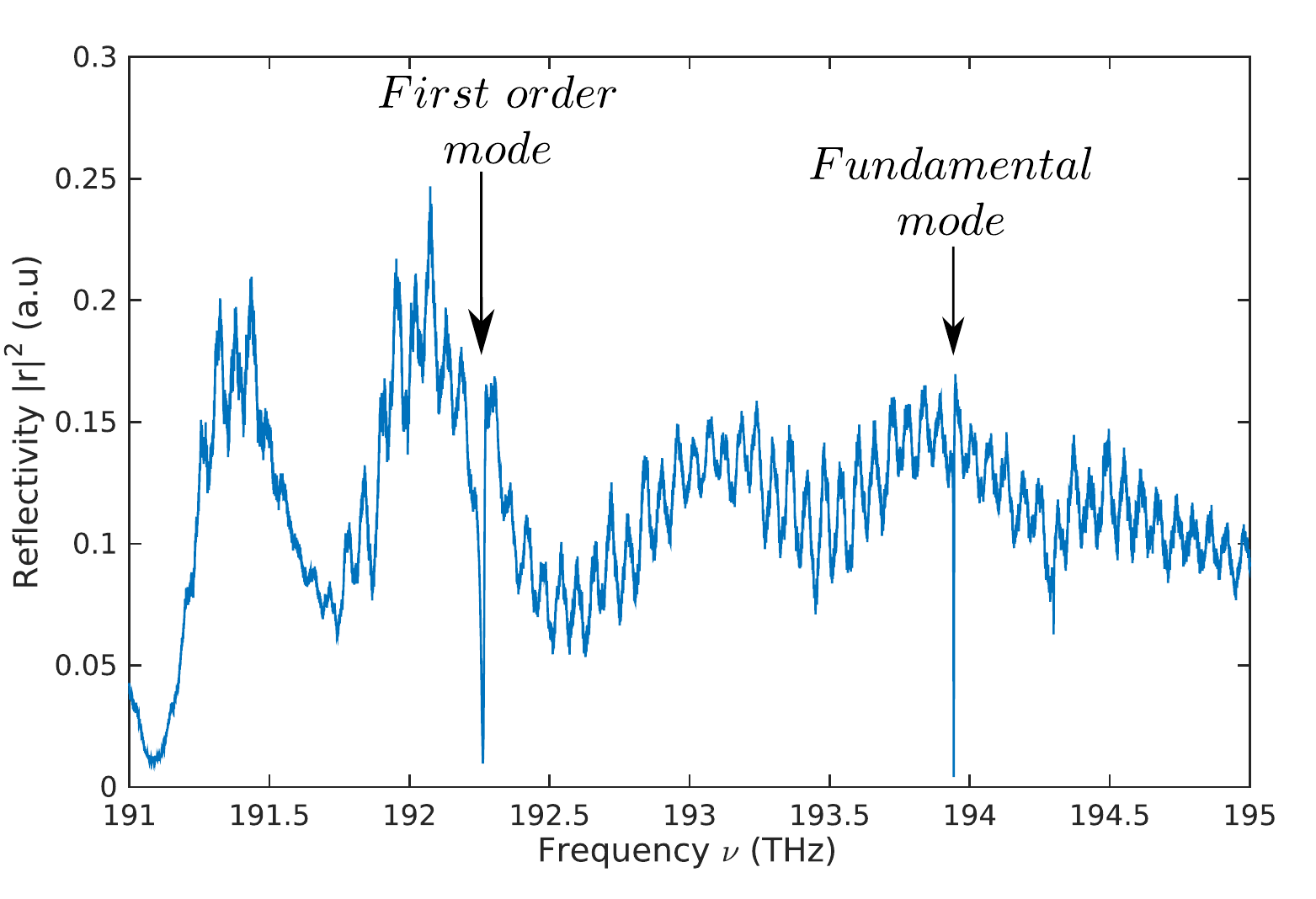}
\caption{Optical spectrum of a cavity with $Q_L=120\ 000$ for the fundamental mode. The fundamental mode and the first order mode can be seen at respectively 194 THz and 192.25 THz}
\label{fig:whole_spectrum}
\end{figure}
The spectrum on Fig.\ref{fig:whole_spectrum} is recorded by Optical Coherence Tomography method \cite{combrie2017compact}. Two resonances can be seen on this spectrum : the resonance with the highest frequency (the fundamental mode) is around 194 THz and the next resonance (the first order mode) is around 192.25 THz. Interferences can also be seen in the reflection spectrum : the low frequency signature is attributed to the interference between the input of the waveguide and the fiber facet whereas the high frequency feature is linked to an interference between the input of the waveguide and the input of the photonic crystal. Further analysis of the reflection spectrum of such a cavity can be found in \cite{Lian_2017}.
\appendix
\section*{Mechanical spectrum}
\begin{figure}[ht!]
\centering
\includegraphics[width=0.9\columnwidth]{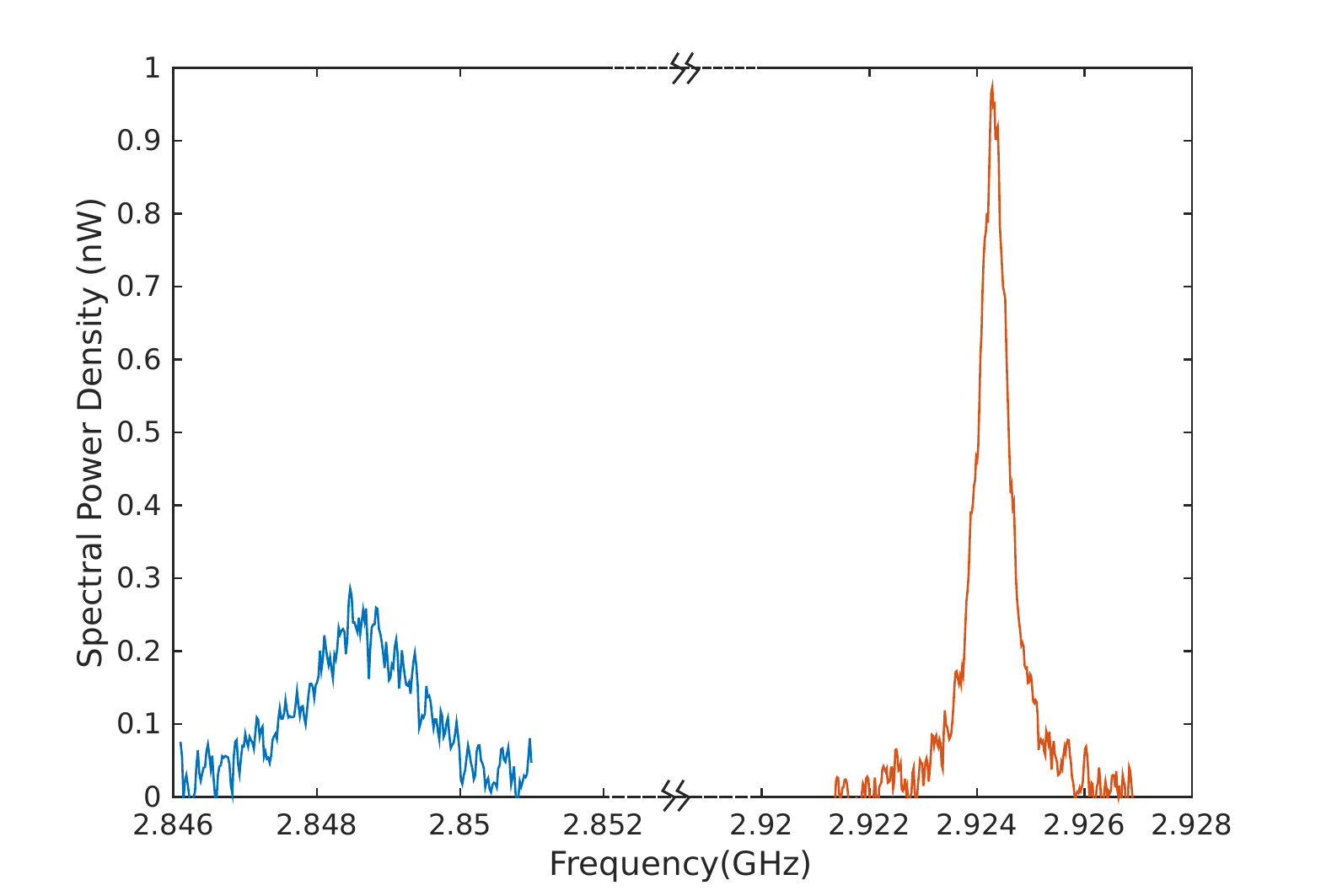}
\caption{Mechanical spectrum of the bichromatic optomechanical crystal. The fundamental mode (orange) and the second order mode (blue) can be seen at respectively 2.92 GHz and 2.85 GHz}
\label{fig:mech_whole_spectrum}
\end{figure}
As can be seen on Fig.\ref{fig:mech_whole_spectrum}, the fundamental mode at 2.92 GHz is indeed the mode with the highest mechanical frequency. The first order mode cannot be seen as it has an odd symmetry, whereas the fundamental optical mode has an even symmetry. According to calculation, $g_0$ between the fundamental optical mode and the second order mechanical mode is equal to $g_0/2\pi=23 kHz$ which is much smaller than the $g_0/2\pi=380 kHz$ for the coupling between the fundamental optical and mechanical modes.
\appendix
\section*{Extraction of the Complex Amplitude Spectrum}
The interferogram $s(\nu)$ which is measured with the OCT system is related to the complex amplitude of the optical field from the sample $\tilde{E}(\nu)=r(\nu)E_r$ through: $s=\tilde{E}E_r^* + c.c.=r \left| E_r \right| ^2 + c.c.$, where $E_r$ is the reference field. The transfer function (here the complex reflectivity) $r(\nu)$ can be retrieved using the Hilbert transform, as shown for instance in ~\cite{gottesman2004new,gottesman2010time} to extract a complex spectrum from the time interferogram $r(t)$ measured by continuously changing the length of one of the arms of an unbalanced Michelson interferometer and a partially coherent light source. Here, the Hilbert transform is applied to a signal in the frequency domain, but the procedure is formally identical. This is achieved by taking the inverse Fourier transform $S$ of the interferogram $s$ and then calculating: $R(t) = S(t) + sign(t)S(t)$ and finally by Fourier transforming again to obtain $r(\nu)$.\\
From $r(\nu)$, the intrinsic losses $\Gamma_0$ and the coupling losses $\gamma$ can be deduced \cite{combrie2017compact} :
\begin{equation}
	r(\nu)=\frac{2\pi\nu-z}{2\pi\nu-p}
\end{equation}
with $z$ and $p$ equal to :
\begin{eqnarray}
	z&=2\pi\nu_0+i(\Gamma_0-\gamma)/2\\
	p&=2\pi\nu_0+i(\Gamma_0+\gamma)/2
\end{eqnarray}
\appendix
\section*{Thermo Optic induced spectral shift}
For a linear evolution of the resonance frequency with the on-chip power, the time-domain Coupled Mode Theory \cite{Joannopoulos_2003} yields the following equation for the true value of the detuning :
\begin{equation}
\Delta'=\Delta - \alpha P_{L}\frac{1}{1+\left[\frac{\Delta^\prime}{\kappa}\right]^{2}}
\end{equation}
where $\Delta'=2\pi(\nu_{L}-\nu_{0}')$ and $\Delta=2\pi(\nu_{L}-\nu_{0})$, $\nu_{0}'$ being the current resonance frequency and $\nu_{0}$ the "cold" cavity resonance frequency. \\
From eq. (5) of \cite{Carmon2004}, the maximum temperature change the system can undergo occurs when $\nu_L=\nu_0^\prime$ or equivalently, when $\Delta=\Delta\nu_{bist}$, $\Delta^\prime=0$ . Therefore, the true detuning can be written as a function of $\kappa$:
\begin{equation}
	\Delta^\prime=\Delta - \frac{\Delta\nu_{bist}}{1+(4\pi\Delta')^{2}/\kappa^2}
\end{equation}
The above equation is therefore solved to obtain the real detuning.\\
\appendix
\section*{Estimation of the on-chip power}
Optical power is coupled into the photonic crystal cavity using a fiber-collimator and a microscope objective. When the pump is out of resonance, these two optical component are the main sources of losses. 
Therefore, the on-chip power is estimated by taking into account losses coming from the collimator and the objective :
\begin{eqnarray}
	P_{on-chip}=\alpha_{c}\alpha_{m}P_{input}&\\
	P_{reflected}=\alpha_{m}\alpha_{c}P_{on-chip}&
\end{eqnarray}
where $\alpha_c$ corresponds to the loss due to the fiber-collimator and $\alpha_m$ represents the loss due to the microscope objective.
Therefore, the on-chip power is found using the formula below : 
\begin{equation}
	P_{on-chip}=\sqrt{P_{reflected}P_{input}}
\end{equation}
\appendix
\section*{Measurement of the vacuum OM coupling}
%
The optical source is a Keysight tuneable laser. The laser is then modulated by a MPZ LN 10 phase modulator from Photline Technologies. After coupling into the cavity, the reflected light is detected by an Optilab APR-10-M APD photodetector and analyzed by a Rhode and Schwarz FSV 40 Electrical Spectrum Analyser (ESA). The fiber link is entirely polarization maintaining. \\
The measurement of $g_{0}$ is carried out according to the method described in ~\cite{gorodetksy2010determination}.  The optomechanical coupling corresponds to the optical frequency shift resulting from the displacement of the mechanical resonator, therefore the method consist in measuring the power density spectrum of the frequency shift $S_{\nu\nu}(f)$ at thermal equilibrium, where the average amplitude of the thermal mechanical fluctuation is known and corresponds to $n_{th}=\frac{k_{B}T}{\hbar\Omega_{m}}$ phonons.\\ 
The vacuum coupling constant is therefore (by definition):
\begin{equation}
g_0^2 = \frac{\int{S_{\nu\nu}(f)df}}{n_{th}}
\end{equation}
where the integral\footnote{here we consider the single-sided spectrum} is computed about the mechanical resonance $f_m$.
The unknown transduction coefficient relating $S_{\nu\nu}$ to the measured electric power spectrum $S$ is determined using a calibration tone generated by phase modulator which is inserted in the input path between the light source and the cavity.\\
\begin{equation}
\int{S_{\nu\nu}(f)df}= \frac{\int{S(f)df}}{\int{S_{cal}(f)df}}\Phi_0^2\frac{(\pi f_m)^2}{4}
\end{equation}
where $S_{cal}$ is the spectral power density in the phase modulation peak and $\phi_{0}=\pi\frac{V_{cal}}{V_{\pi}}$, with $V_{\pi}=6.11$ V.\\
As the ESA measures the electrical power $\tilde{S}$ within the selected resolution bandwidth $RBW$, it follows that $\int{S_{cal}(f)df}=\tilde{S}_{cal}$ as the calibration tone is spectrally narrower than $RBW$. In contrast, the spectrum of the frequency fluctuations of the OM oscillator is broader and, following \cite{gorodetksy2010determination}, its integral is evaluated from the fitted Lorentzian lineshape with FWHM $\Gamma_m$ as: $\int{S(f)df}=\max(\tilde{S})\Gamma_m/RBW$. This leads to the known formula:
\begin{equation}
g_{0}^{2}=\frac{\max(\tilde{S})}{S_{cal}}\frac{\Gamma_{m}}{n_{th}}\phi_{0}^{2} \frac{\pi^2 f_m^{2}}{4RBW}
\end{equation}
The experiment is carried out by setting the calibration tone away from the resonance but still close enough such that the transduction function can still be considered constant.
\appendix
\section*{Parameters used in the model}
\begin{center}
\begin{tabular}{|p{1.75cm}|p{4.1cm}|p{0.85cm}|p{1.5cm}|}
    \hline
    \multirow{3}{2em}{Optical properties}& Coupled quality factor & $Q$ & $30\ 000$  \\
                               & Intrinsic quality factor &$Q_0$ & $200\ 000$\\
                               & Resonance Frequency (THz) & $\nu_0$ & $193.79$ \\
    \hline
    \multirow{4}{2em}{Mechanical properties}&Mechanical frequency(GHz)&$f_m$& $2.92$\\
    									    &Quality factor& $Q_m$& 2300\\
    									    &Zero point fluctuation (fm)& $x_{ZPF}$ & $1.6$\\
    									    &Effective mass(fg)&$m_{eff}$&$1.07$\\
    \hline
    \multirow{2}{2em}{Thermal properties}&Relaxation time $(\mu s)$ & $\tau_{th}$&18\\
    									 &Linear thermal expansion ($10^{-6}K$)&$\alpha$&5.3\\
    									 &Thermomechanical force (nN)&$F_{th}$&$0.6$\\
    \hline
    \multicolumn{2}{|p{5.95cm}|}{Frequency shift per displacement ($Hz/m^{-1}$)} & G & $1.51.10^{21}$\\
    \hline
    \multicolumn{2}{|p{5.95cm}|}{Optomechanical coupling ($kHz$)} & $g_{0}$ & $2\pi.385$\\
    \hline
\end{tabular}
\end{center}
\section*{Influence of photothermal forces on anti-damping and optical spring}
\begin{figure}[ht!]
	\centering
	\includegraphics[width=0.9\columnwidth]{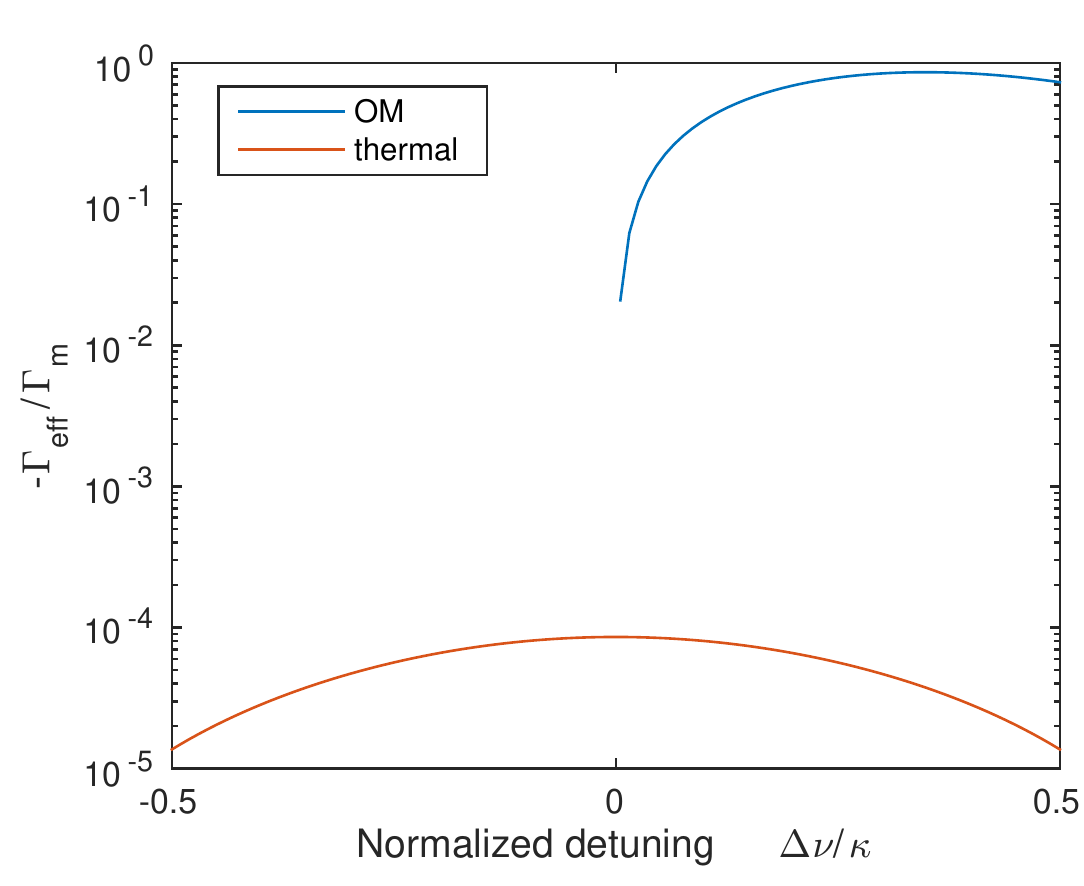}
	\caption{Comparison of the photothermal contribution and the contribution of the radiation pressure to antidamping as a function of normalized detuning}
	\label{fig:thermo}
\end{figure}

To quantify the influence of photothermal forces, we use the model developed in \cite{Guha2017}, which takes into account the evolution of temperature in the OMC :
\begin{eqnarray}
    &m_{eff}\ddot{x}+\Gamma_{m}m_{eff}\dot{x}+\omega_{m}^2x=\hbar G\left|a\right|^2+F_{th}\\
    &\dot{a}+\left(\frac{\kappa}{2}-i\Delta-iGx-\frac{\omega_{cav}}{n}\frac{dn}{dT}\right)a-\kappa_{ex}a_{in}=0\\
    &\frac{d\Delta T}{dt}=-\frac{\Delta T}{\tau_{th}}+\frac{\Gamma_{th}\left|a\right|^2}{\tau_{th}}
\end{eqnarray}
where $\Gamma_{th}=R_{th}\hbar\omega_L\kappa_{abs}$. $\tau_{th}$ is the thermal relaxation time, which is found by numerical simulation. $F_{th}$ is the photothermal force, found by considering the influence of a linear expansion of the OMC due to one photon absorbed. By linearizing around an equilibrium point, we find the expressions for the optical spring and anti-damping as a function of normalized detuning $x=\frac{\Delta}{\kappa}$ :
\begin{equation}
\begin{split}
    &\Gamma_{eff}=\frac{G}{m_{eff}\Omega\hbar\nu_{L}}\frac{\kappa_{ex}P}{\left[x^2+1/4\right]\kappa^2}\left[ \frac{1}{2\kappa}\left[\hbar G+\frac{F_{th}}{1+(\Omega\tau_{th})^2}\right]\right.\\
    & \left. \left[\frac{1}{1/4+(x+\Omega/\kappa)^2}-\frac{1}{1/4+(x-\Omega/\kappa)^2}\right]+\frac{F_{th}\Omega\tau_{th}}{1+(\Omega\tau_{th})^2}\right.\\
    &\left. \left[\frac{x-\Omega/\kappa}{1/4+(x-\Omega/\kappa)^2}-\frac{x+\Omega/\kappa}{1/4+(x+\Omega/\kappa)^2}\right]\right]
\end{split}
\end{equation}
\begin{equation}
\begin{split}
	&\delta\Omega_{eff}=\frac{G}{m_{eff}\Omega}\frac{\kappa_{ex}P}{\left[x^2+1/4\right]\kappa^2}\frac{1}{\hbar\Omega_{L}}\left[\frac{1}{2\kappa}\left[\hbar G+\frac{F_{th}}{1+(\Omega\tau_{th})^2}\right]\right.\\
	&\left[\frac{x+\Omega/\kappa}{1/4+(x+\Omega/\kappa)^2}+\frac{x-\Omega/\kappa}{1/4+(x-\Omega/\kappa)^2}\right]+\\
    &\left.\frac{F_{th}\Omega\tau_{th}}{1+(\Omega\tau_{th})^2}\frac{1}{\kappa}\left[\frac{1/2}{1/4+(x-\Omega/\kappa)^2}-\frac{1/2}{1/4+(x+\Omega/\kappa)^2}\right]\right]
\end{split}
\end{equation}
From these equations, one can surmise that the influence of photothermal forces is negligible when the relaxation time is slow compared to the oscillation dynamics. Indeed, when plotting the contribution of photothermal forces to antidamping and comparing it to the antidamping due to radiation pressure (Fig \ref{fig:thermo}), a difference of 4 orders of magnitude is clear between the two contributions.
\appendix
\section*{Measurement of the RF resonance}
The RF spectra are fitted using the Voigt lineshape. This function is defined as the convolution of a Lorentzian lineshape $\mathcal{L}(x)=\gamma\pi^{-1} (x^2+\gamma^2)^{-1}$ and a Gaussian broadening function $\mathcal{G}(x)=\exp(-x^2/2\sigma^2)/\sqrt{2\pi}\sigma$, namely:
\begin{equation}
V(x;\gamma,\sigma)=\int{\mathcal{G}(x;\sigma)\mathcal{L}(x-x^\prime;\gamma)dx^\prime}
\end{equation} 
 
The Voigt function is calculated efficiently through the Faddeeva function $w(z)$, (implemented in \footnote{\textcolor{blue}{http://ab-initio.mit.edu/wiki/index.php/Faddeeva\_Package}, written by S. Johnson.}), through the relations: $V(x;\gamma,\sigma)=\mathcal{R}[w(z)]/\sigma\sqrt{2\pi}$ and $z=(x+\imath\gamma)/\sigma\sqrt{2}$.

The fit is taken considering data points above the noise level estimated at -55 dB below the peak. \\
The integrated power is obtained by integrating the raw spectra $P_{RF}=\int{S}/RBW$, the resolution bandwidth, as above, unless the linewidth is narrower than the instrument resolution, where the peak level is taken instead.
\appendix
\section*{Measurement of the phase noise}
Phase noise is measured through the heterodyne technique \cite{howe1981} using a Frequency Synthesizer as  local oscillator at $f_{LO}$. First, the optical signal extracted from the cavity is mixed with a continuous wave strong optical carrier. The optical signal obtained after mixing is sent to a balanced photodetector (Discovery Semiconductors). The electrical signal is amplified using a 20 dB Mini Circuits amplifier then further amplified by another 40 dB (Femto Amplifier) before the mixer. The low frequency signal $v(t)$ is digitized with a 12bit real time sampling oscilloscope (Lecroy HDO), sampling time $1.25\times10^6$ samples/s, with $N=2.5\times10^6$ samples. Then, the signal $v_n=v(n\Delta t)$ is processed as in \cite{maxin2014}
First $v_n$ is multiplied by $\exp(-2\pi\imath f_0 \Delta t n)$ and Fourier transformed using FFT (denoted as $\mathcal{F})$. Then the low-frequency part of the spectrum ($|f|<BW$) is transformed back in the time domain, which gives the analytic signal $v_a$ around the carrier frequency $f_0=f_{OM}-f_{LO}$. The phase is obtained by taking the argument of each sample $\phi_n=arg(v_a(t_n))$.
Then the power spectral density of the phase is evaluated within a certain spectral band $f\in[f_i , f_{i+1}]$ using the standard procedure. This defines a time span $1/2 f_i$ long enough to resolve $f_i$. Consequently $\phi_n$ is distributed in $N_i$ consecutive windows $\Delta W_j$ with duration $1/2 f_i$ and containing $M_i$ samples, such that $M_i N_i=N $. In each time window, the non stationary contributions (trend and average) are removed and then a suitable window function (Hanning  $h_n$) is applied to the signal $\tilde{\phi}_{n,j,i}$, before Fourier transform (FFT). Finally, the power spectra $S_{j,i}(f_{k,i})= |\mathcal{F}(\tilde{\phi}_{n,j,i}h_n)|^2$ are averaged over the windows, being , $f_{k,i}\in[f_i , f_{i+1}]$ .  More precisely:

\begin{equation}
\mathcal{L}({f_{k,i}) \approx  \frac{\Delta t}{N_i M_i^2} \frac{ \sum_{j=1}^{N_i} {S_{j,i} (f_{k,i}) }} 
{\sum_n^{M_i}{h_n^2}} }
\end{equation}

\begin{thebibliography}{51}%
\makeatletter
\providecommand \@ifxundefined [1]{%
 \@ifx{#1\undefined}
}%
\providecommand \@ifnum [1]{%
 \ifnum #1\expandafter \@firstoftwo
 \else \expandafter \@secondoftwo
 \fi
}%
\providecommand \@ifx [1]{%
 \ifx #1\expandafter \@firstoftwo
 \else \expandafter \@secondoftwo
 \fi
}%
\providecommand \natexlab [1]{#1}%
\providecommand \enquote  [1]{``#1''}%
\providecommand \bibnamefont  [1]{#1}%
\providecommand \bibfnamefont [1]{#1}%
\providecommand \citenamefont [1]{#1}%
\providecommand \href@noop [0]{\@secondoftwo}%
\providecommand \href [0]{\begingroup \@sanitize@url \@href}%
\providecommand \@href[1]{\@@startlink{#1}\@@href}%
\providecommand \@@href[1]{\endgroup#1\@@endlink}%
\providecommand \@sanitize@url [0]{\catcode `\\12\catcode `\$12\catcode
  `\&12\catcode `\#12\catcode `\^12\catcode `\_12\catcode `\%12\relax}%
\providecommand \@@startlink[1]{}%
\providecommand \@@endlink[0]{}%
\providecommand \url  [0]{\begingroup\@sanitize@url \@url }%
\providecommand \@url [1]{\endgroup\@href {#1}{\urlprefix }}%
\providecommand \urlprefix  [0]{URL }%
\providecommand \Eprint [0]{\href }%
\providecommand \doibase [0]{http://dx.doi.org/}%
\providecommand \selectlanguage [0]{\@gobble}%
\providecommand \bibinfo  [0]{\@secondoftwo}%
\providecommand \bibfield  [0]{\@secondoftwo}%
\providecommand \translation [1]{[#1]}%
\providecommand \BibitemOpen [0]{}%
\providecommand \bibitemStop [0]{}%
\providecommand \bibitemNoStop [0]{.\EOS\space}%
\providecommand \EOS [0]{\spacefactor3000\relax}%
\providecommand \BibitemShut  [1]{\csname bibitem#1\endcsname}%
\let\auto@bib@innerbib\@empty
\bibitem [{\citenamefont {Kippenberg}\ and\ \citenamefont
  {Vahala}(2008)}]{Kippenberg2008}%
  \BibitemOpen
  \bibfield  {author} {\bibinfo {author} {\bibfnamefont {T.~J.}\ \bibnamefont
  {Kippenberg}}\ and\ \bibinfo {author} {\bibfnamefont {K.~J.}\ \bibnamefont
  {Vahala}},\ }\href {\doibase 10.1126/science.1156032} {\bibfield  {journal}
  {\bibinfo  {journal} {Science}\ }\textbf {\bibinfo {volume} {321}},\ \bibinfo
  {pages} {1172} (\bibinfo {year} {2008})}\BibitemShut {NoStop}%
\bibitem [{\citenamefont {Riedinger}\ \emph {et~al.}(2018)\citenamefont
  {Riedinger}, \citenamefont {Wallucks}, \citenamefont {Marinkovi{\'c}},
  \citenamefont {L{\"o}schnauer}, \citenamefont {Aspelmeyer}, \citenamefont
  {Hong},\ and\ \citenamefont {Gr{\"o}blacher}}]{Riedinger_2018}%
  \BibitemOpen
  \bibfield  {author} {\bibinfo {author} {\bibfnamefont {R.}~\bibnamefont
  {Riedinger}}, \bibinfo {author} {\bibfnamefont {A.}~\bibnamefont {Wallucks}},
  \bibinfo {author} {\bibfnamefont {I.}~\bibnamefont {Marinkovi{\'c}}},
  \bibinfo {author} {\bibfnamefont {C.}~\bibnamefont {L{\"o}schnauer}},
  \bibinfo {author} {\bibfnamefont {M.}~\bibnamefont {Aspelmeyer}}, \bibinfo
  {author} {\bibfnamefont {S.}~\bibnamefont {Hong}}, \ and\ \bibinfo {author}
  {\bibfnamefont {S.}~\bibnamefont {Gr{\"o}blacher}},\ }\href@noop {}
  {\bibfield  {journal} {\bibinfo  {journal} {Nature}\ }\textbf {\bibinfo
  {volume} {556}},\ \bibinfo {pages} {473} (\bibinfo {year}
  {2018})}\BibitemShut {NoStop}%
\bibitem [{\citenamefont {Cohen}\ \emph {et~al.}(2015)\citenamefont {Cohen},
  \citenamefont {Meenehan}, \citenamefont {MacCabe}, \citenamefont
  {Gr{\"o}blacher}, \citenamefont {Safavi-Naeini}, \citenamefont {Marsili},
  \citenamefont {Shaw},\ and\ \citenamefont {Painter}}]{Cohen_2015}%
  \BibitemOpen
  \bibfield  {author} {\bibinfo {author} {\bibfnamefont {J.~D.}\ \bibnamefont
  {Cohen}}, \bibinfo {author} {\bibfnamefont {S.~M.}\ \bibnamefont {Meenehan}},
  \bibinfo {author} {\bibfnamefont {G.~S.}\ \bibnamefont {MacCabe}}, \bibinfo
  {author} {\bibfnamefont {S.}~\bibnamefont {Gr{\"o}blacher}}, \bibinfo
  {author} {\bibfnamefont {A.~H.}\ \bibnamefont {Safavi-Naeini}}, \bibinfo
  {author} {\bibfnamefont {F.}~\bibnamefont {Marsili}}, \bibinfo {author}
  {\bibfnamefont {M.~D.}\ \bibnamefont {Shaw}}, \ and\ \bibinfo {author}
  {\bibfnamefont {O.}~\bibnamefont {Painter}},\ }\href@noop {} {\bibfield
  {journal} {\bibinfo  {journal} {Nature}\ }\textbf {\bibinfo {volume} {520}},\
  \bibinfo {pages} {522} (\bibinfo {year} {2015})}\BibitemShut {NoStop}%
\bibitem [{\citenamefont {Purdy}\ \emph {et~al.}(2017)\citenamefont {Purdy},
  \citenamefont {Grutter}, \citenamefont {Srinivasan},\ and\ \citenamefont
  {Taylor}}]{Purdy_2017}%
  \BibitemOpen
  \bibfield  {author} {\bibinfo {author} {\bibfnamefont {T.~P.}\ \bibnamefont
  {Purdy}}, \bibinfo {author} {\bibfnamefont {K.~E.}\ \bibnamefont {Grutter}},
  \bibinfo {author} {\bibfnamefont {K.}~\bibnamefont {Srinivasan}}, \ and\
  \bibinfo {author} {\bibfnamefont {J.~M.}\ \bibnamefont {Taylor}},\ }\href
  {\doibase 10.1126/science.aag1407} {\bibfield  {journal} {\bibinfo  {journal}
  {Science}\ }\textbf {\bibinfo {volume} {356}},\ \bibinfo {pages} {1265}
  (\bibinfo {year} {2017})}\BibitemShut {NoStop}%
\bibitem [{\citenamefont {Krause}\ \emph {et~al.}(2012)\citenamefont {Krause},
  \citenamefont {Winger}, \citenamefont {Blasius}, \citenamefont {Lin},\ and\
  \citenamefont {Painter}}]{Krause2012_accelero}%
  \BibitemOpen
  \bibfield  {author} {\bibinfo {author} {\bibfnamefont {A.~G.}\ \bibnamefont
  {Krause}}, \bibinfo {author} {\bibfnamefont {M.}~\bibnamefont {Winger}},
  \bibinfo {author} {\bibfnamefont {T.~D.}\ \bibnamefont {Blasius}}, \bibinfo
  {author} {\bibfnamefont {Q.}~\bibnamefont {Lin}}, \ and\ \bibinfo {author}
  {\bibfnamefont {O.}~\bibnamefont {Painter}},\ }\href@noop {} {\bibfield
  {journal} {\bibinfo  {journal} {Nature Photonics}\ }\textbf {\bibinfo
  {volume} {6}},\ \bibinfo {pages} {768} (\bibinfo {year} {2012})}\BibitemShut
  {NoStop}%
\bibitem [{\citenamefont {Balram}\ \emph {et~al.}(2016)\citenamefont {Balram},
  \citenamefont {Davan\c{c}o}, \citenamefont {Song},\ and\ \citenamefont
  {Srinivasan}}]{Balram2016}%
  \BibitemOpen
  \bibfield  {author} {\bibinfo {author} {\bibfnamefont {K.~C.}\ \bibnamefont
  {Balram}}, \bibinfo {author} {\bibfnamefont {M.~I.}\ \bibnamefont
  {Davan\c{c}o}}, \bibinfo {author} {\bibfnamefont {J.~D.}\ \bibnamefont
  {Song}}, \ and\ \bibinfo {author} {\bibfnamefont {K.}~\bibnamefont
  {Srinivasan}},\ }\href
  {http://www.nature.com/doifinder/10.1038/nphoton.2016.46} {\bibfield
  {journal} {\bibinfo  {journal} {Nature Photonics}\ ,\ \bibinfo {pages} {346}}
  (\bibinfo {year} {2016})}\BibitemShut {NoStop}%
\bibitem [{\citenamefont {Massel}\ \emph {et~al.}(2011)\citenamefont {Massel},
  \citenamefont {Heikkil{\"a}}, \citenamefont {Pirkkalainen}, \citenamefont
  {Cho}, \citenamefont {Saloniemi}, \citenamefont {Hakonen},\ and\
  \citenamefont {Sillanp{\"a}{\"a}}}]{massel_2011}%
  \BibitemOpen
  \bibfield  {author} {\bibinfo {author} {\bibfnamefont {F.}~\bibnamefont
  {Massel}}, \bibinfo {author} {\bibfnamefont {T.}~\bibnamefont
  {Heikkil{\"a}}}, \bibinfo {author} {\bibfnamefont {J.-M.}\ \bibnamefont
  {Pirkkalainen}}, \bibinfo {author} {\bibfnamefont {S.-U.}\ \bibnamefont
  {Cho}}, \bibinfo {author} {\bibfnamefont {H.}~\bibnamefont {Saloniemi}},
  \bibinfo {author} {\bibfnamefont {P.~J.}\ \bibnamefont {Hakonen}}, \ and\
  \bibinfo {author} {\bibfnamefont {M.~A.}\ \bibnamefont {Sillanp{\"a}{\"a}}},\
  }\href@noop {} {\bibfield  {journal} {\bibinfo  {journal} {Nature}\ }\textbf
  {\bibinfo {volume} {480}},\ \bibinfo {pages} {351} (\bibinfo {year}
  {2011})}\BibitemShut {NoStop}%
\bibitem [{\citenamefont {Hossein-Zadeh}\ and\ \citenamefont
  {Vahala}(2010)}]{Hossein-Zadeh2010}%
  \BibitemOpen
  \bibfield  {author} {\bibinfo {author} {\bibfnamefont {M.}~\bibnamefont
  {Hossein-Zadeh}}\ and\ \bibinfo {author} {\bibfnamefont {K.~J.}\ \bibnamefont
  {Vahala}},\ }\href@noop {} {\bibfield  {journal} {\bibinfo  {journal} {IEEE
  Journal of Selected Topics in Quantum Electronics}\ }\textbf {\bibinfo
  {volume} {16}},\ \bibinfo {pages} {276} (\bibinfo {year} {2010})}\BibitemShut
  {NoStop}%
\bibitem [{\citenamefont {Eichenfield}\ \emph {et~al.}(2009)\citenamefont
  {Eichenfield}, \citenamefont {Chan}, \citenamefont {Camacho}, \citenamefont
  {Vahala},\ and\ \citenamefont {Painter}}]{Eichenfield2009}%
  \BibitemOpen
  \bibfield  {author} {\bibinfo {author} {\bibfnamefont {M.}~\bibnamefont
  {Eichenfield}}, \bibinfo {author} {\bibfnamefont {J.}~\bibnamefont {Chan}},
  \bibinfo {author} {\bibfnamefont {R.~M.}\ \bibnamefont {Camacho}}, \bibinfo
  {author} {\bibfnamefont {K.~J.}\ \bibnamefont {Vahala}}, \ and\ \bibinfo
  {author} {\bibfnamefont {O.}~\bibnamefont {Painter}},\ }\href {\doibase
  10.1038/nature08524} {\bibfield  {journal} {\bibinfo  {journal} {Nature}\
  }\textbf {\bibinfo {volume} {462}},\ \bibinfo {pages} {78} (\bibinfo {year}
  {2009})}\BibitemShut {NoStop}%
\bibitem [{\citenamefont {Enzian}\ \emph {et~al.}(2019)\citenamefont {Enzian},
  \citenamefont {Szczykulska}, \citenamefont {Silver}, \citenamefont {Bino},
  \citenamefont {Zhang}, \citenamefont {Walmsley}, \citenamefont {Del'Haye},\
  and\ \citenamefont {Vanner}}]{Enzian_2019}%
  \BibitemOpen
  \bibfield  {author} {\bibinfo {author} {\bibfnamefont {G.}~\bibnamefont
  {Enzian}}, \bibinfo {author} {\bibfnamefont {M.}~\bibnamefont {Szczykulska}},
  \bibinfo {author} {\bibfnamefont {J.}~\bibnamefont {Silver}}, \bibinfo
  {author} {\bibfnamefont {L.~D.}\ \bibnamefont {Bino}}, \bibinfo {author}
  {\bibfnamefont {S.}~\bibnamefont {Zhang}}, \bibinfo {author} {\bibfnamefont
  {I.~A.}\ \bibnamefont {Walmsley}}, \bibinfo {author} {\bibfnamefont
  {P.}~\bibnamefont {Del'Haye}}, \ and\ \bibinfo {author} {\bibfnamefont
  {M.~R.}\ \bibnamefont {Vanner}},\ }\href@noop {} {\bibfield  {journal}
  {\bibinfo  {journal} {Optica}\ }\textbf {\bibinfo {volume} {6}},\ \bibinfo
  {pages} {7} (\bibinfo {year} {2019})}\BibitemShut {NoStop}%
\bibitem [{\citenamefont {Davan\c{c}o}\ \emph {et~al.}(2014)\citenamefont
  {Davan\c{c}o}, \citenamefont {Ates}, \citenamefont {Liu},\ and\ \citenamefont
  {Srinivasan}}]{davanco_si3n4_2014}%
  \BibitemOpen
  \bibfield  {author} {\bibinfo {author} {\bibfnamefont {M.}~\bibnamefont
  {Davan\c{c}o}}, \bibinfo {author} {\bibfnamefont {S.}~\bibnamefont {Ates}},
  \bibinfo {author} {\bibfnamefont {Y.}~\bibnamefont {Liu}}, \ and\ \bibinfo
  {author} {\bibfnamefont {K.}~\bibnamefont {Srinivasan}},\ }\href {\doibase
  10.1063/1.4858975} {\bibfield  {journal} {\bibinfo  {journal} {Applied
  Physics Letters}\ }\textbf {\bibinfo {volume} {104}},\ \bibinfo {pages}
  {041101} (\bibinfo {year} {2014})}\BibitemShut {NoStop}%
\bibitem [{\citenamefont {Burek}\ \emph {et~al.}(2016)\citenamefont {Burek},
  \citenamefont {Cohen}, \citenamefont {Meenehan}, \citenamefont {El-Sawah},
  \citenamefont {Chia}, \citenamefont {Ruelle}, \citenamefont {Meesala},
  \citenamefont {Rochman}, \citenamefont {Atikian}, \citenamefont {Markham},
  \citenamefont {Twitchen}, \citenamefont {Lukin}, \citenamefont {Painter},\
  and\ \citenamefont {Lon\v{c}ar}}]{Burek2016}%
  \BibitemOpen
  \bibfield  {author} {\bibinfo {author} {\bibfnamefont {M.~J.}\ \bibnamefont
  {Burek}}, \bibinfo {author} {\bibfnamefont {J.~D.}\ \bibnamefont {Cohen}},
  \bibinfo {author} {\bibfnamefont {S.~M.}\ \bibnamefont {Meenehan}}, \bibinfo
  {author} {\bibfnamefont {N.}~\bibnamefont {El-Sawah}}, \bibinfo {author}
  {\bibfnamefont {C.}~\bibnamefont {Chia}}, \bibinfo {author} {\bibfnamefont
  {T.}~\bibnamefont {Ruelle}}, \bibinfo {author} {\bibfnamefont
  {S.}~\bibnamefont {Meesala}}, \bibinfo {author} {\bibfnamefont
  {J.}~\bibnamefont {Rochman}}, \bibinfo {author} {\bibfnamefont {H.~A.}\
  \bibnamefont {Atikian}}, \bibinfo {author} {\bibfnamefont {M.}~\bibnamefont
  {Markham}}, \bibinfo {author} {\bibfnamefont {D.~J.}\ \bibnamefont
  {Twitchen}}, \bibinfo {author} {\bibfnamefont {M.~D.}\ \bibnamefont {Lukin}},
  \bibinfo {author} {\bibfnamefont {O.}~\bibnamefont {Painter}}, \ and\
  \bibinfo {author} {\bibfnamefont {M.}~\bibnamefont {Lon\v{c}ar}},\
  }\href@noop {} {\bibfield  {journal} {\bibinfo  {journal} {Optica}\ }\textbf
  {\bibinfo {volume} {3}},\ \bibinfo {pages} {1404} (\bibinfo {year}
  {2016})}\BibitemShut {NoStop}%
\bibitem [{\citenamefont {Mitchell}\ \emph {et~al.}(2016)\citenamefont
  {Mitchell}, \citenamefont {Khanaliloo}, \citenamefont {Lake}, \citenamefont
  {Masuda}, \citenamefont {Hadden},\ and\ \citenamefont
  {Barclay}}]{OM_single_diamond_crystal}%
  \BibitemOpen
  \bibfield  {author} {\bibinfo {author} {\bibfnamefont {M.}~\bibnamefont
  {Mitchell}}, \bibinfo {author} {\bibfnamefont {B.}~\bibnamefont
  {Khanaliloo}}, \bibinfo {author} {\bibfnamefont {D.~P.}\ \bibnamefont
  {Lake}}, \bibinfo {author} {\bibfnamefont {T.}~\bibnamefont {Masuda}},
  \bibinfo {author} {\bibfnamefont {J.~P.}\ \bibnamefont {Hadden}}, \ and\
  \bibinfo {author} {\bibfnamefont {P.~E.}\ \bibnamefont {Barclay}},\ }\href
  {\doibase 10.1364/OPTICA.3.000963} {\bibfield  {journal} {\bibinfo  {journal}
  {Optica}\ }\textbf {\bibinfo {volume} {3}},\ \bibinfo {pages} {963} (\bibinfo
  {year} {2016})}\BibitemShut {NoStop}%
\bibitem [{\citenamefont {Midolo}\ \emph {et~al.}(2018)\citenamefont {Midolo},
  \citenamefont {Schliesser},\ and\ \citenamefont {Fiore}}]{Midolo_2018}%
  \BibitemOpen
  \bibfield  {author} {\bibinfo {author} {\bibfnamefont {L.}~\bibnamefont
  {Midolo}}, \bibinfo {author} {\bibfnamefont {A.}~\bibnamefont {Schliesser}},
  \ and\ \bibinfo {author} {\bibfnamefont {A.}~\bibnamefont {Fiore}},\
  }\href@noop {} {\bibfield  {journal} {\bibinfo  {journal} {Nature
  nanotechnology}\ }\textbf {\bibinfo {volume} {13}},\ \bibinfo {pages} {11}
  (\bibinfo {year} {2018})}\BibitemShut {NoStop}%
\bibitem [{\citenamefont {Barzanjeh}\ \emph {et~al.}(2015)\citenamefont
  {Barzanjeh}, \citenamefont {Guha}, \citenamefont {Weedbrook}, \citenamefont
  {Vitali}, \citenamefont {Shapiro},\ and\ \citenamefont
  {Pirandola}}]{Barzanjeh2015}%
  \BibitemOpen
  \bibfield  {author} {\bibinfo {author} {\bibfnamefont {S.}~\bibnamefont
  {Barzanjeh}}, \bibinfo {author} {\bibfnamefont {S.}~\bibnamefont {Guha}},
  \bibinfo {author} {\bibfnamefont {C.}~\bibnamefont {Weedbrook}}, \bibinfo
  {author} {\bibfnamefont {D.}~\bibnamefont {Vitali}}, \bibinfo {author}
  {\bibfnamefont {J.~H.}\ \bibnamefont {Shapiro}}, \ and\ \bibinfo {author}
  {\bibfnamefont {S.}~\bibnamefont {Pirandola}},\ }\href {\doibase
  10.1103/PhysRevLett.114.080503} {\bibfield  {journal} {\bibinfo  {journal}
  {Phys. Rev. Lett.}\ }\textbf {\bibinfo {volume} {114}},\ \bibinfo {pages}
  {080503} (\bibinfo {year} {2015})}\BibitemShut {NoStop}%
\bibitem [{\citenamefont {Schneider}\ \emph {et~al.}(2018)\citenamefont
  {Schneider}, \citenamefont {Baumgartner}, \citenamefont {H\"{o}nl},
  \citenamefont {Welter}, \citenamefont {Hahn}, \citenamefont {Wilson},
  \citenamefont {Czornomaz},\ and\ \citenamefont {Seidler}}]{Schneider2018}%
  \BibitemOpen
  \bibfield  {author} {\bibinfo {author} {\bibfnamefont {K.}~\bibnamefont
  {Schneider}}, \bibinfo {author} {\bibfnamefont {Y.}~\bibnamefont
  {Baumgartner}}, \bibinfo {author} {\bibfnamefont {S.}~\bibnamefont
  {H\"{o}nl}}, \bibinfo {author} {\bibfnamefont {P.}~\bibnamefont {Welter}},
  \bibinfo {author} {\bibfnamefont {H.}~\bibnamefont {Hahn}}, \bibinfo {author}
  {\bibfnamefont {D.~J.}\ \bibnamefont {Wilson}}, \bibinfo {author}
  {\bibfnamefont {L.}~\bibnamefont {Czornomaz}}, \ and\ \bibinfo {author}
  {\bibfnamefont {P.}~\bibnamefont {Seidler}},\ }\href@noop {} {\bibfield
  {journal} {\bibinfo  {journal} {arXiv:1812.00631}\ } (\bibinfo {year}
  {2018})}\BibitemShut {NoStop}%
\bibitem [{\citenamefont {Mitchell}\ \emph {et~al.}(2014)\citenamefont
  {Mitchell}, \citenamefont {Hryciw},\ and\ \citenamefont
  {Barclay}}]{GaP_microdisk}%
  \BibitemOpen
  \bibfield  {author} {\bibinfo {author} {\bibfnamefont {M.}~\bibnamefont
  {Mitchell}}, \bibinfo {author} {\bibfnamefont {A.~C.}\ \bibnamefont
  {Hryciw}}, \ and\ \bibinfo {author} {\bibfnamefont {P.~E.}\ \bibnamefont
  {Barclay}},\ }\href@noop {} {\bibfield  {journal} {\bibinfo  {journal}
  {Applied Physics Letters}\ }\textbf {\bibinfo {volume} {104}},\ \bibinfo
  {pages} {141104} (\bibinfo {year} {2014})}\BibitemShut {NoStop}%
\bibitem [{\citenamefont {Bochmann}\ \emph {et~al.}(2013)\citenamefont
  {Bochmann}, \citenamefont {Vainsencher}, \citenamefont {Awschalom},\ and\
  \citenamefont {Cleland}}]{Bochmann2013}%
  \BibitemOpen
  \bibfield  {author} {\bibinfo {author} {\bibfnamefont {J.}~\bibnamefont
  {Bochmann}}, \bibinfo {author} {\bibfnamefont {A.}~\bibnamefont
  {Vainsencher}}, \bibinfo {author} {\bibfnamefont {D.~D.}\ \bibnamefont
  {Awschalom}}, \ and\ \bibinfo {author} {\bibfnamefont {A.~N.}\ \bibnamefont
  {Cleland}},\ }\href {\doibase 10.1038/nphys2748} {\bibfield  {journal}
  {\bibinfo  {journal} {Nature Physics}\ }\textbf {\bibinfo {volume} {9}},\
  \bibinfo {pages} {712} (\bibinfo {year} {2013})}\BibitemShut {NoStop}%
\bibitem [{\citenamefont {Combri{\'e}}\ \emph {et~al.}(2009)\citenamefont
  {Combri{\'e}}, \citenamefont {Tran}, \citenamefont {De~Rossi}, \citenamefont
  {Husko},\ and\ \citenamefont {Colman}}]{combrie2009high}%
  \BibitemOpen
  \bibfield  {author} {\bibinfo {author} {\bibfnamefont {S.}~\bibnamefont
  {Combri{\'e}}}, \bibinfo {author} {\bibfnamefont {Q.~V.}\ \bibnamefont
  {Tran}}, \bibinfo {author} {\bibfnamefont {A.}~\bibnamefont {De~Rossi}},
  \bibinfo {author} {\bibfnamefont {C.}~\bibnamefont {Husko}}, \ and\ \bibinfo
  {author} {\bibfnamefont {P.}~\bibnamefont {Colman}},\ }\href@noop {}
  {\bibfield  {journal} {\bibinfo  {journal} {Appl. Phys. Lett.}\ }\textbf
  {\bibinfo {volume} {95}},\ \bibinfo {pages} {221108} (\bibinfo {year}
  {2009})}\BibitemShut {NoStop}%
\bibitem [{\citenamefont {Colman}\ \emph {et~al.}(2010)\citenamefont {Colman},
  \citenamefont {Husko}, \citenamefont {Combri{\'e}}, \citenamefont {Sagnes},
  \citenamefont {Wong},\ and\ \citenamefont {De~Rossi}}]{colman2010temporal}%
  \BibitemOpen
  \bibfield  {author} {\bibinfo {author} {\bibfnamefont {P.}~\bibnamefont
  {Colman}}, \bibinfo {author} {\bibfnamefont {C.}~\bibnamefont {Husko}},
  \bibinfo {author} {\bibfnamefont {S.}~\bibnamefont {Combri{\'e}}}, \bibinfo
  {author} {\bibfnamefont {I.}~\bibnamefont {Sagnes}}, \bibinfo {author}
  {\bibfnamefont {C.~W.}\ \bibnamefont {Wong}}, \ and\ \bibinfo {author}
  {\bibfnamefont {A.}~\bibnamefont {De~Rossi}},\ }\href@noop {} {\bibfield
  {journal} {\bibinfo  {journal} {Nature Photonics}\ }\textbf {\bibinfo
  {volume} {4}},\ \bibinfo {pages} {862} (\bibinfo {year} {2010})}\BibitemShut
  {NoStop}%
\bibitem [{\citenamefont {B\"{u}ckle}\ \emph {et~al.}(2018)\citenamefont
  {B\"{u}ckle}, \citenamefont {Hauber}, \citenamefont {Cole}, \citenamefont
  {G\"{a}rtner}, \citenamefont {Zeimer}, \citenamefont {Grenzer},\ and\
  \citenamefont {Weig}}]{Buckle_2018}%
  \BibitemOpen
  \bibfield  {author} {\bibinfo {author} {\bibfnamefont {M.}~\bibnamefont
  {B\"{u}ckle}}, \bibinfo {author} {\bibfnamefont {V.~C.}\ \bibnamefont
  {Hauber}}, \bibinfo {author} {\bibfnamefont {G.~D.}\ \bibnamefont {Cole}},
  \bibinfo {author} {\bibfnamefont {C.}~\bibnamefont {G\"{a}rtner}}, \bibinfo
  {author} {\bibfnamefont {U.}~\bibnamefont {Zeimer}}, \bibinfo {author}
  {\bibfnamefont {J.}~\bibnamefont {Grenzer}}, \ and\ \bibinfo {author}
  {\bibfnamefont {E.~M.}\ \bibnamefont {Weig}},\ }\href@noop {} {\bibfield
  {journal} {\bibinfo  {journal} {Applied Physics Letters}\ }\textbf {\bibinfo
  {volume} {113}},\ \bibinfo {pages} {201903} (\bibinfo {year}
  {2018})}\BibitemShut {NoStop}%
\bibitem [{\citenamefont {Cole}\ \emph {et~al.}(2014)\citenamefont {Cole},
  \citenamefont {Yu}, \citenamefont {G\"{a}rtner}, \citenamefont {Siquans},
  \citenamefont {Moghadas~Nia}, \citenamefont {Schm\"{o}le}, \citenamefont
  {Hoelscher-Obermaier}, \citenamefont {Purdy}, \citenamefont {Wieczorek},
  \citenamefont {Regal},\ and\ \citenamefont {Aspelmeyer}}]{Cole2014_GaInP}%
  \BibitemOpen
  \bibfield  {author} {\bibinfo {author} {\bibfnamefont {G.~D.}\ \bibnamefont
  {Cole}}, \bibinfo {author} {\bibfnamefont {P.-L.}\ \bibnamefont {Yu}},
  \bibinfo {author} {\bibfnamefont {C.}~\bibnamefont {G\"{a}rtner}}, \bibinfo
  {author} {\bibfnamefont {K.}~\bibnamefont {Siquans}}, \bibinfo {author}
  {\bibfnamefont {R.}~\bibnamefont {Moghadas~Nia}}, \bibinfo {author}
  {\bibfnamefont {J.}~\bibnamefont {Schm\"{o}le}}, \bibinfo {author}
  {\bibfnamefont {J.}~\bibnamefont {Hoelscher-Obermaier}}, \bibinfo {author}
  {\bibfnamefont {T.~P.}\ \bibnamefont {Purdy}}, \bibinfo {author}
  {\bibfnamefont {W.}~\bibnamefont {Wieczorek}}, \bibinfo {author}
  {\bibfnamefont {C.~A.}\ \bibnamefont {Regal}}, \ and\ \bibinfo {author}
  {\bibfnamefont {M.}~\bibnamefont {Aspelmeyer}},\ }\href@noop {} {\bibfield
  {journal} {\bibinfo  {journal} {Applied Physics Letters}\ }\textbf {\bibinfo
  {volume} {104}},\ \bibinfo {pages} {201908} (\bibinfo {year}
  {2014})}\BibitemShut {NoStop}%
\bibitem [{\citenamefont {Guha}\ \emph {et~al.}(2017)\citenamefont {Guha},
  \citenamefont {Mariani}, \citenamefont {Lema\^itre}, \citenamefont
  {Combri\'e}, \citenamefont {Leo},\ and\ \citenamefont {Favero}}]{Guha2017}%
  \BibitemOpen
  \bibfield  {author} {\bibinfo {author} {\bibfnamefont {B.}~\bibnamefont
  {Guha}}, \bibinfo {author} {\bibfnamefont {S.}~\bibnamefont {Mariani}},
  \bibinfo {author} {\bibfnamefont {A.}~\bibnamefont {Lema\^itre}}, \bibinfo
  {author} {\bibfnamefont {S.}~\bibnamefont {Combri\'e}}, \bibinfo {author}
  {\bibfnamefont {G.}~\bibnamefont {Leo}}, \ and\ \bibinfo {author}
  {\bibfnamefont {I.}~\bibnamefont {Favero}},\ }\href {\doibase
  10.1364/OE.25.024639} {\bibfield  {journal} {\bibinfo  {journal} {Optics
  Express}\ }\textbf {\bibinfo {volume} {25}},\ \bibinfo {pages} {24639}
  (\bibinfo {year} {2017})}\BibitemShut {NoStop}%
\bibitem [{\citenamefont {Combri{\'e}}\ \emph {et~al.}(2017)\citenamefont
  {Combri{\'e}}, \citenamefont {Lehoucq}, \citenamefont {Moille}, \citenamefont
  {Martin},\ and\ \citenamefont {De~Rossi}}]{combrie2017compact}%
  \BibitemOpen
  \bibfield  {author} {\bibinfo {author} {\bibfnamefont {S.}~\bibnamefont
  {Combri{\'e}}}, \bibinfo {author} {\bibfnamefont {G.}~\bibnamefont
  {Lehoucq}}, \bibinfo {author} {\bibfnamefont {G.}~\bibnamefont {Moille}},
  \bibinfo {author} {\bibfnamefont {A.}~\bibnamefont {Martin}}, \ and\ \bibinfo
  {author} {\bibfnamefont {A.}~\bibnamefont {De~Rossi}},\ }\href@noop {}
  {\bibfield  {journal} {\bibinfo  {journal} {Laser and Photonics Reviews}\
  }\textbf {\bibinfo {volume} {11}},\ \bibinfo {pages} {1700099} (\bibinfo
  {year} {2017})}\BibitemShut {NoStop}%
\bibitem [{\citenamefont {Dodane}\ \emph {et~al.}(2018)\citenamefont {Dodane},
  \citenamefont {Bourderionnet}, \citenamefont {Combri{\'e}},\ and\
  \citenamefont {De~Rossi}}]{delphin_2018}%
  \BibitemOpen
  \bibfield  {author} {\bibinfo {author} {\bibfnamefont {D.}~\bibnamefont
  {Dodane}}, \bibinfo {author} {\bibfnamefont {J.}~\bibnamefont
  {Bourderionnet}}, \bibinfo {author} {\bibfnamefont {S.}~\bibnamefont
  {Combri{\'e}}}, \ and\ \bibinfo {author} {\bibfnamefont {A.}~\bibnamefont
  {De~Rossi}},\ }\href@noop {} {\bibfield  {journal} {\bibinfo  {journal}
  {Optics Express}\ }\textbf {\bibinfo {volume} {26}},\ \bibinfo {pages}
  {20868} (\bibinfo {year} {2018})}\BibitemShut {NoStop}%
\bibitem [{\citenamefont {Safavi-Naeini}\ \emph {et~al.}(2011)\citenamefont
  {Safavi-Naeini}, \citenamefont {Alegre}, \citenamefont {Chan}, \citenamefont
  {Eichenfield}, \citenamefont {Winger}, \citenamefont {Lin}, \citenamefont
  {Hill}, \citenamefont {Chang},\ and\ \citenamefont
  {Painter}}]{Safavi-Naeini2011}%
  \BibitemOpen
  \bibfield  {author} {\bibinfo {author} {\bibfnamefont {A.~H.}\ \bibnamefont
  {Safavi-Naeini}}, \bibinfo {author} {\bibfnamefont {T.~P.~M.}\ \bibnamefont
  {Alegre}}, \bibinfo {author} {\bibfnamefont {J.}~\bibnamefont {Chan}},
  \bibinfo {author} {\bibfnamefont {M.}~\bibnamefont {Eichenfield}}, \bibinfo
  {author} {\bibfnamefont {M.}~\bibnamefont {Winger}}, \bibinfo {author}
  {\bibfnamefont {Q.}~\bibnamefont {Lin}}, \bibinfo {author} {\bibfnamefont
  {J.~T.}\ \bibnamefont {Hill}}, \bibinfo {author} {\bibfnamefont {D.~E.}\
  \bibnamefont {Chang}}, \ and\ \bibinfo {author} {\bibfnamefont
  {O.}~\bibnamefont {Painter}},\ }\href {\doibase 10.1038/nature09933}
  {\bibfield  {journal} {\bibinfo  {journal} {Nature}\ }\textbf {\bibinfo
  {volume} {472}},\ \bibinfo {pages} {69} (\bibinfo {year} {2011})}\BibitemShut
  {NoStop}%
\bibitem [{\citenamefont {Gomis-Bresco}\ \emph {et~al.}(2014)\citenamefont
  {Gomis-Bresco}, \citenamefont {Navarro-Urrios}, \citenamefont {Oudich},
  \citenamefont {El-Jallal}, \citenamefont {Griol}, \citenamefont {Puerto},
  \citenamefont {Chavez}, \citenamefont {Pennec}, \citenamefont
  {Djafari-Rouhani}, \citenamefont {Alzina} \emph {et~al.}}]{gomis2014one}%
  \BibitemOpen
  \bibfield  {author} {\bibinfo {author} {\bibfnamefont {J.}~\bibnamefont
  {Gomis-Bresco}}, \bibinfo {author} {\bibfnamefont {D.}~\bibnamefont
  {Navarro-Urrios}}, \bibinfo {author} {\bibfnamefont {M.}~\bibnamefont
  {Oudich}}, \bibinfo {author} {\bibfnamefont {S.}~\bibnamefont {El-Jallal}},
  \bibinfo {author} {\bibfnamefont {A.}~\bibnamefont {Griol}}, \bibinfo
  {author} {\bibfnamefont {D.}~\bibnamefont {Puerto}}, \bibinfo {author}
  {\bibfnamefont {E.}~\bibnamefont {Chavez}}, \bibinfo {author} {\bibfnamefont
  {Y.}~\bibnamefont {Pennec}}, \bibinfo {author} {\bibfnamefont
  {B.}~\bibnamefont {Djafari-Rouhani}}, \bibinfo {author} {\bibfnamefont
  {F.}~\bibnamefont {Alzina}},  \emph {et~al.},\ }\href@noop {} {\bibfield
  {journal} {\bibinfo  {journal} {Nature Communications}\ }\textbf {\bibinfo
  {volume} {5}},\ \bibinfo {pages} {4452} (\bibinfo {year} {2014})}\BibitemShut
  {NoStop}%
\bibitem [{\citenamefont {Song}\ \emph {et~al.}()\citenamefont {Song},
  \citenamefont {Noda}, \citenamefont {Asano},\ and\ \citenamefont
  {Akahane}}]{Song_2005}%
  \BibitemOpen
  \bibfield  {author} {\bibinfo {author} {\bibfnamefont {B.-S.}\ \bibnamefont
  {Song}}, \bibinfo {author} {\bibfnamefont {S.}~\bibnamefont {Noda}}, \bibinfo
  {author} {\bibfnamefont {T.}~\bibnamefont {Asano}}, \ and\ \bibinfo {author}
  {\bibfnamefont {Y.}~\bibnamefont {Akahane}},\ }\href@noop {} {\bibfield
  {journal} {\bibinfo  {journal} {Nature Materials}\ }\textbf {\bibinfo
  {volume} {4}},\ \bibinfo {pages} {207}}\BibitemShut {NoStop}%
\bibitem [{\citenamefont {Alpeggiani}\ \emph {et~al.}(2015)\citenamefont
  {Alpeggiani}, \citenamefont {Andreani},\ and\ \citenamefont
  {Gerace}}]{alpeggiani_effective_2015}%
  \BibitemOpen
  \bibfield  {author} {\bibinfo {author} {\bibfnamefont {F.}~\bibnamefont
  {Alpeggiani}}, \bibinfo {author} {\bibfnamefont {L.~C.}\ \bibnamefont
  {Andreani}}, \ and\ \bibinfo {author} {\bibfnamefont {D.}~\bibnamefont
  {Gerace}},\ }\href {\doibase 10.1063/1.4938395} {\bibfield  {journal}
  {\bibinfo  {journal} {Applied Physics Letters}\ }\textbf {\bibinfo {volume}
  {107}},\ \bibinfo {pages} {261110} (\bibinfo {year} {2015})}\BibitemShut
  {NoStop}%
\bibitem [{\citenamefont {Baker}\ \emph {et~al.}(2014)\citenamefont {Baker},
  \citenamefont {Hease}, \citenamefont {Nguyen}, \citenamefont {Andronico},
  \citenamefont {Ducci}, \citenamefont {Leo},\ and\ \citenamefont
  {Favero}}]{Baker_2014}%
  \BibitemOpen
  \bibfield  {author} {\bibinfo {author} {\bibfnamefont {C.}~\bibnamefont
  {Baker}}, \bibinfo {author} {\bibfnamefont {W.}~\bibnamefont {Hease}},
  \bibinfo {author} {\bibfnamefont {D.-T.}\ \bibnamefont {Nguyen}}, \bibinfo
  {author} {\bibfnamefont {A.}~\bibnamefont {Andronico}}, \bibinfo {author}
  {\bibfnamefont {S.}~\bibnamefont {Ducci}}, \bibinfo {author} {\bibfnamefont
  {G.}~\bibnamefont {Leo}}, \ and\ \bibinfo {author} {\bibfnamefont
  {I.}~\bibnamefont {Favero}},\ }\href@noop {} {\bibfield  {journal} {\bibinfo
  {journal} {Opt. Express}\ }\textbf {\bibinfo {volume} {22}},\ \bibinfo
  {pages} {14072} (\bibinfo {year} {2014})}\BibitemShut {NoStop}%
\bibitem [{\citenamefont {Tran}\ \emph {et~al.}(2009)\citenamefont {Tran},
  \citenamefont {Combri{\'e}}, \citenamefont {Colman},\ and\ \citenamefont
  {De~Rossi}}]{tran2009photonic}%
  \BibitemOpen
  \bibfield  {author} {\bibinfo {author} {\bibfnamefont {Q.~V.}\ \bibnamefont
  {Tran}}, \bibinfo {author} {\bibfnamefont {S.}~\bibnamefont {Combri{\'e}}},
  \bibinfo {author} {\bibfnamefont {P.}~\bibnamefont {Colman}}, \ and\ \bibinfo
  {author} {\bibfnamefont {A.}~\bibnamefont {De~Rossi}},\ }\href@noop {}
  {\bibfield  {journal} {\bibinfo  {journal} {Appl. Phys. Lett.}\ }\textbf
  {\bibinfo {volume} {95}},\ \bibinfo {pages} {061105} (\bibinfo {year}
  {2009})}\BibitemShut {NoStop}%
\bibitem [{\citenamefont {Carmon}\ \emph {et~al.}(2004)\citenamefont {Carmon},
  \citenamefont {Yang},\ and\ \citenamefont {Vahala}}]{Carmon2004}%
  \BibitemOpen
  \bibfield  {author} {\bibinfo {author} {\bibfnamefont {T.}~\bibnamefont
  {Carmon}}, \bibinfo {author} {\bibfnamefont {L.}~\bibnamefont {Yang}}, \ and\
  \bibinfo {author} {\bibfnamefont {K.~J.}\ \bibnamefont {Vahala}},\
  }\href@noop {} {\bibfield  {journal} {\bibinfo  {journal} {Optics Express}\
  }\textbf {\bibinfo {volume} {12}},\ \bibinfo {pages} {4742} (\bibinfo {year}
  {2004})}\BibitemShut {NoStop}%
\bibitem [{\citenamefont {Martin}\ \emph {et~al.}(2017)\citenamefont {Martin},
  \citenamefont {Sanchez}, \citenamefont {Combri\'{e}}, \citenamefont
  {de~Rossi},\ and\ \citenamefont {Raineri}}]{martin2017gainp}%
  \BibitemOpen
  \bibfield  {author} {\bibinfo {author} {\bibfnamefont {A.}~\bibnamefont
  {Martin}}, \bibinfo {author} {\bibfnamefont {D.}~\bibnamefont {Sanchez}},
  \bibinfo {author} {\bibfnamefont {S.}~\bibnamefont {Combri\'{e}}}, \bibinfo
  {author} {\bibfnamefont {A.}~\bibnamefont {de~Rossi}}, \ and\ \bibinfo
  {author} {\bibfnamefont {F.}~\bibnamefont {Raineri}},\ }\href {\doibase
  10.1364/OL.42.000599} {\bibfield  {journal} {\bibinfo  {journal} {Opt.
  Lett.}\ }\textbf {\bibinfo {volume} {42}},\ \bibinfo {pages} {599} (\bibinfo
  {year} {2017})}\BibitemShut {NoStop}%
\bibitem [{\citenamefont {Gorodetksy}\ \emph {et~al.}(2010)\citenamefont
  {Gorodetksy}, \citenamefont {Schliesser}, \citenamefont {Anetsberger},
  \citenamefont {Deleglise},\ and\ \citenamefont
  {Kippenberg}}]{gorodetksy2010determination}%
  \BibitemOpen
  \bibfield  {author} {\bibinfo {author} {\bibfnamefont {M.}~\bibnamefont
  {Gorodetksy}}, \bibinfo {author} {\bibfnamefont {A.}~\bibnamefont
  {Schliesser}}, \bibinfo {author} {\bibfnamefont {G.}~\bibnamefont
  {Anetsberger}}, \bibinfo {author} {\bibfnamefont {S.}~\bibnamefont
  {Deleglise}}, \ and\ \bibinfo {author} {\bibfnamefont {T.~J.}\ \bibnamefont
  {Kippenberg}},\ }\href {\doibase 10.1364/OE.18.023236} {\bibfield  {journal}
  {\bibinfo  {journal} {Optics express}\ }\textbf {\bibinfo {volume} {18}},\
  \bibinfo {pages} {23236} (\bibinfo {year} {2010})}\BibitemShut {NoStop}%
\bibitem [{\citenamefont {Aspelmeyer}\ \emph {et~al.}(2014)\citenamefont
  {Aspelmeyer}, \citenamefont {Kippenberg},\ and\ \citenamefont
  {Marquardt}}]{Aspelmeyer2014}%
  \BibitemOpen
  \bibfield  {author} {\bibinfo {author} {\bibfnamefont {M.}~\bibnamefont
  {Aspelmeyer}}, \bibinfo {author} {\bibfnamefont {T.~J.}\ \bibnamefont
  {Kippenberg}}, \ and\ \bibinfo {author} {\bibfnamefont {F.}~\bibnamefont
  {Marquardt}},\ }\href {\doibase 10.1103/RevModPhys.86.1391} {\bibfield
  {journal} {\bibinfo  {journal} {Reviews of Modern Physics}\ }\textbf
  {\bibinfo {volume} {86}},\ \bibinfo {pages} {1391} (\bibinfo {year}
  {2014})}\BibitemShut {NoStop}%
\bibitem [{\citenamefont {Vahala}(2008)}]{Vahala2008backactionlimit}%
  \BibitemOpen
  \bibfield  {author} {\bibinfo {author} {\bibfnamefont {K.~J.}\ \bibnamefont
  {Vahala}},\ }\href {\doibase 10.1103/PhysRevA.78.023832} {\bibfield
  {journal} {\bibinfo  {journal} {Phys. Rev. A}\ }\textbf {\bibinfo {volume}
  {78}},\ \bibinfo {pages} {023832} (\bibinfo {year} {2008})}\BibitemShut
  {NoStop}%
\bibitem [{\citenamefont {Tallur}\ \emph {et~al.}(2011)\citenamefont {Tallur},
  \citenamefont {Sridaran},\ and\ \citenamefont {Bhave}}]{Tallur_2011}%
  \BibitemOpen
  \bibfield  {author} {\bibinfo {author} {\bibfnamefont {S.}~\bibnamefont
  {Tallur}}, \bibinfo {author} {\bibfnamefont {S.}~\bibnamefont {Sridaran}}, \
  and\ \bibinfo {author} {\bibfnamefont {S.~A.}\ \bibnamefont {Bhave}},\ }\href
  {\doibase 10.1364/OE.19.024522} {\bibfield  {journal} {\bibinfo  {journal}
  {Optics Express}\ }\textbf {\bibinfo {volume} {19}},\ \bibinfo {pages}
  {24522} (\bibinfo {year} {2011})}\BibitemShut {NoStop}%
\bibitem [{\citenamefont {Sridaran}\ and\ \citenamefont
  {Bhave}(2012)}]{Sridaran_2012}%
  \BibitemOpen
  \bibfield  {author} {\bibinfo {author} {\bibfnamefont {S.}~\bibnamefont
  {Sridaran}}\ and\ \bibinfo {author} {\bibfnamefont {S.~A.}\ \bibnamefont
  {Bhave}},\ }in\ \href@noop {} {\emph {\bibinfo {booktitle} {2012 IEEE 25th
  International Conference on Micro Electro Mechanical Systems (MEMS)}}}\
  (\bibinfo  {publisher} {IEEE},\ \bibinfo {year} {2012})\ pp.\ \bibinfo
  {pages} {664--667}\BibitemShut {NoStop}%
\bibitem [{\citenamefont {Tsvirkun}\ \emph {et~al.}(2015)\citenamefont
  {Tsvirkun}, \citenamefont {Surrente}, \citenamefont {Raineri}, \citenamefont
  {Beaudoin}, \citenamefont {Raj}, \citenamefont {Sagnes}, \citenamefont
  {Robert-Philip},\ and\ \citenamefont {Braive}}]{Integrated1652}%
  \BibitemOpen
  \bibfield  {author} {\bibinfo {author} {\bibfnamefont {V.}~\bibnamefont
  {Tsvirkun}}, \bibinfo {author} {\bibfnamefont {A.}~\bibnamefont {Surrente}},
  \bibinfo {author} {\bibfnamefont {F.}~\bibnamefont {Raineri}}, \bibinfo
  {author} {\bibfnamefont {G.}~\bibnamefont {Beaudoin}}, \bibinfo {author}
  {\bibfnamefont {R.}~\bibnamefont {Raj}}, \bibinfo {author} {\bibfnamefont
  {I.}~\bibnamefont {Sagnes}}, \bibinfo {author} {\bibfnamefont
  {I.}~\bibnamefont {Robert-Philip}}, \ and\ \bibinfo {author} {\bibfnamefont
  {R.}~\bibnamefont {Braive}},\ }\href@noop {} {\bibfield  {journal} {\bibinfo
  {journal} {Scientific Reports}\ }\textbf {\bibinfo {volume} {5}},\ \bibinfo
  {pages} {16526} (\bibinfo {year} {2015})}\BibitemShut {NoStop}%
\bibitem [{\citenamefont {Matsko}\ \emph {et~al.}(2011)\citenamefont {Matsko},
  \citenamefont {Savchenkov}, \citenamefont {Ilchenko}, \citenamefont
  {Seidel},\ and\ \citenamefont {Maleki}}]{Matsko2011_selfref}%
  \BibitemOpen
  \bibfield  {author} {\bibinfo {author} {\bibfnamefont {A.~B.}\ \bibnamefont
  {Matsko}}, \bibinfo {author} {\bibfnamefont {A.~A.}\ \bibnamefont
  {Savchenkov}}, \bibinfo {author} {\bibfnamefont {V.~S.}\ \bibnamefont
  {Ilchenko}}, \bibinfo {author} {\bibfnamefont {D.}~\bibnamefont {Seidel}}, \
  and\ \bibinfo {author} {\bibfnamefont {L.}~\bibnamefont {Maleki}},\ }\href
  {\doibase 10.1103/PhysRevA.83.021801(R)} {\bibfield  {journal} {\bibinfo
  {journal} {Phys. Rev. A}\ }\textbf {\bibinfo {volume} {83}},\ \bibinfo
  {pages} {021801(R)} (\bibinfo {year} {2011})}\BibitemShut {NoStop}%
\bibitem [{\citenamefont {Ghadimi}\ \emph {et~al.}(2018)\citenamefont
  {Ghadimi}, \citenamefont {Fedorov}, \citenamefont {Engelsen}, \citenamefont
  {Bereyhi}, \citenamefont {Schilling}, \citenamefont {Wilson},\ and\
  \citenamefont {Kippenberg}}]{Ghadimi_2018}%
  \BibitemOpen
  \bibfield  {author} {\bibinfo {author} {\bibfnamefont {A.~H.}\ \bibnamefont
  {Ghadimi}}, \bibinfo {author} {\bibfnamefont {S.~A.}\ \bibnamefont
  {Fedorov}}, \bibinfo {author} {\bibfnamefont {N.~J.}\ \bibnamefont
  {Engelsen}}, \bibinfo {author} {\bibfnamefont {M.~J.}\ \bibnamefont
  {Bereyhi}}, \bibinfo {author} {\bibfnamefont {R.}~\bibnamefont {Schilling}},
  \bibinfo {author} {\bibfnamefont {D.~J.}\ \bibnamefont {Wilson}}, \ and\
  \bibinfo {author} {\bibfnamefont {T.~J.}\ \bibnamefont {Kippenberg}},\
  }\href@noop {} {\bibfield  {journal} {\bibinfo  {journal} {Science}\ }\textbf
  {\bibinfo {volume} {360}},\ \bibinfo {pages} {764} (\bibinfo {year}
  {2018})}\BibitemShut {NoStop}%
\bibitem [{\citenamefont {Navarro-Urrios}\ \emph {et~al.}(2017)\citenamefont
  {Navarro-Urrios}, \citenamefont {Capuj}, \citenamefont {Colombano},
  \citenamefont {Garcia}, \citenamefont {Sledzinska}, \citenamefont {Alzina},
  \citenamefont {Griol}, \citenamefont {Alejandro},\ and\ \citenamefont
  {Sotomayor-Torres}}]{Sotomayor_non_linear_dynamics}%
  \BibitemOpen
  \bibfield  {author} {\bibinfo {author} {\bibfnamefont {D.}~\bibnamefont
  {Navarro-Urrios}}, \bibinfo {author} {\bibfnamefont {N.~E.}\ \bibnamefont
  {Capuj}}, \bibinfo {author} {\bibfnamefont {M.~F.}\ \bibnamefont
  {Colombano}}, \bibinfo {author} {\bibfnamefont {P.~D.}\ \bibnamefont
  {Garcia}}, \bibinfo {author} {\bibfnamefont {M.}~\bibnamefont {Sledzinska}},
  \bibinfo {author} {\bibfnamefont {F.}~\bibnamefont {Alzina}}, \bibinfo
  {author} {\bibfnamefont {A.}~\bibnamefont {Griol}}, \bibinfo {author}
  {\bibfnamefont {M.}~\bibnamefont {Alejandro}}, \ and\ \bibinfo {author}
  {\bibfnamefont {C.~M.}\ \bibnamefont {Sotomayor-Torres}},\ }\href@noop {}
  {\bibfield  {journal} {\bibinfo  {journal} {Nature Communications}\ }\textbf
  {\bibinfo {volume} {8}},\ \bibinfo {pages} {14965} (\bibinfo {year}
  {2017})}\BibitemShut {NoStop}%
\bibitem [{\citenamefont {Heinrich}\ \emph {et~al.}(2011)\citenamefont
  {Heinrich}, \citenamefont {Ludwig}, \citenamefont {Qian}, \citenamefont
  {Kubala},\ and\ \citenamefont {Marquardt}}]{Heinrich2011}%
  \BibitemOpen
  \bibfield  {author} {\bibinfo {author} {\bibfnamefont {G.}~\bibnamefont
  {Heinrich}}, \bibinfo {author} {\bibfnamefont {M.}~\bibnamefont {Ludwig}},
  \bibinfo {author} {\bibfnamefont {J.}~\bibnamefont {Qian}}, \bibinfo {author}
  {\bibfnamefont {B.}~\bibnamefont {Kubala}}, \ and\ \bibinfo {author}
  {\bibfnamefont {F.}~\bibnamefont {Marquardt}},\ }\href@noop {} {\bibfield
  {journal} {\bibinfo  {journal} {Physical Review Letters}\ }\textbf {\bibinfo
  {volume} {107}},\ \bibinfo {pages} {043603} (\bibinfo {year}
  {2011})}\BibitemShut {NoStop}%
\bibitem [{\citenamefont {Balram}\ \emph {et~al.}(2014)\citenamefont {Balram},
  \citenamefont {Davan\c{c}o}, \citenamefont {Lim}, \citenamefont {Song},\ and\
  \citenamefont {Srinivasan}}]{Balram2014}%
  \BibitemOpen
  \bibfield  {author} {\bibinfo {author} {\bibfnamefont {K.~C.}\ \bibnamefont
  {Balram}}, \bibinfo {author} {\bibfnamefont {M.}~\bibnamefont {Davan\c{c}o}},
  \bibinfo {author} {\bibfnamefont {J.~Y.}\ \bibnamefont {Lim}}, \bibinfo
  {author} {\bibfnamefont {J.~D.}\ \bibnamefont {Song}}, \ and\ \bibinfo
  {author} {\bibfnamefont {K.}~\bibnamefont {Srinivasan}},\ }\href {\doibase
  10.1364/OPTICA.1.000414} {\bibfield  {journal} {\bibinfo  {journal} {Optica}\
  }\textbf {\bibinfo {volume} {1}},\ \bibinfo {pages} {414} (\bibinfo {year}
  {2014})}\BibitemShut {NoStop}%
\bibitem [{\citenamefont {Mytsyk}\ \emph {et~al.}(2015)\citenamefont {Mytsyk},
  \citenamefont {Demyanyshyn},\ and\ \citenamefont {Sakharuk}}]{Mytsyk:15}%
  \BibitemOpen
  \bibfield  {author} {\bibinfo {author} {\bibfnamefont {B.~G.}\ \bibnamefont
  {Mytsyk}}, \bibinfo {author} {\bibfnamefont {N.~M.}\ \bibnamefont
  {Demyanyshyn}}, \ and\ \bibinfo {author} {\bibfnamefont {O.~M.}\ \bibnamefont
  {Sakharuk}},\ }\href {\doibase 10.1364/AO.54.008546} {\bibfield  {journal}
  {\bibinfo  {journal} {Appl. Opt.}\ }\textbf {\bibinfo {volume} {54}},\
  \bibinfo {pages} {8546} (\bibinfo {year} {2015})}\BibitemShut {NoStop}%
\bibitem [{\citenamefont {Lian}\ \emph {et~al.}(2017)\citenamefont {Lian},
  \citenamefont {Sokolov}, \citenamefont {Y\"uce}, \citenamefont {Combri\'e},
  \citenamefont {De~Rossi},\ and\ \citenamefont {Mosk}}]{Lian_2017}%
  \BibitemOpen
  \bibfield  {author} {\bibinfo {author} {\bibfnamefont {J.}~\bibnamefont
  {Lian}}, \bibinfo {author} {\bibfnamefont {S.}~\bibnamefont {Sokolov}},
  \bibinfo {author} {\bibfnamefont {E.}~\bibnamefont {Y\"uce}}, \bibinfo
  {author} {\bibfnamefont {S.}~\bibnamefont {Combri\'e}}, \bibinfo {author}
  {\bibfnamefont {A.}~\bibnamefont {De~Rossi}}, \ and\ \bibinfo {author}
  {\bibfnamefont {A.~P.}\ \bibnamefont {Mosk}},\ }\href@noop {} {\bibfield
  {journal} {\bibinfo  {journal} {Phys. Rev. A}\ }\textbf {\bibinfo {volume}
  {96}},\ \bibinfo {pages} {033812} (\bibinfo {year} {2017})}\BibitemShut
  {NoStop}%
\bibitem [{\citenamefont {Gottesman}\ \emph {et~al.}(2004)\citenamefont
  {Gottesman}, \citenamefont {Rao},\ and\ \citenamefont
  {Rabus}}]{gottesman2004new}%
  \BibitemOpen
  \bibfield  {author} {\bibinfo {author} {\bibfnamefont {Y.}~\bibnamefont
  {Gottesman}}, \bibinfo {author} {\bibfnamefont {E.}~\bibnamefont {Rao}}, \
  and\ \bibinfo {author} {\bibfnamefont {D.}~\bibnamefont {Rabus}},\
  }\href@noop {} {\bibfield  {journal} {\bibinfo  {journal} {J. Lightwave
  Tech.}\ }\textbf {\bibinfo {volume} {22}},\ \bibinfo {pages} {1566} (\bibinfo
  {year} {2004})}\BibitemShut {NoStop}%
\bibitem [{\citenamefont {Gottesman}\ \emph {et~al.}(2010)\citenamefont
  {Gottesman}, \citenamefont {Combri{\'e}}, \citenamefont {DeRossi},
  \citenamefont {Talneau}, \citenamefont {Hamel}, \citenamefont {Parini},
  \citenamefont {Gabet}, \citenamefont {Jaouen}, \citenamefont {Benkelfat},\
  and\ \citenamefont {Rao}}]{gottesman2010time}%
  \BibitemOpen
  \bibfield  {author} {\bibinfo {author} {\bibfnamefont {Y.}~\bibnamefont
  {Gottesman}}, \bibinfo {author} {\bibfnamefont {S.}~\bibnamefont
  {Combri{\'e}}}, \bibinfo {author} {\bibfnamefont {A.}~\bibnamefont
  {DeRossi}}, \bibinfo {author} {\bibfnamefont {A.}~\bibnamefont {Talneau}},
  \bibinfo {author} {\bibfnamefont {P.}~\bibnamefont {Hamel}}, \bibinfo
  {author} {\bibfnamefont {A.}~\bibnamefont {Parini}}, \bibinfo {author}
  {\bibfnamefont {R.}~\bibnamefont {Gabet}}, \bibinfo {author} {\bibfnamefont
  {Y.}~\bibnamefont {Jaouen}}, \bibinfo {author} {\bibfnamefont {B.-E.}\
  \bibnamefont {Benkelfat}}, \ and\ \bibinfo {author} {\bibfnamefont {E.~V.}\
  \bibnamefont {Rao}},\ }\href@noop {} {\bibfield  {journal} {\bibinfo
  {journal} {J. Lightwave Tech.}\ }\textbf {\bibinfo {volume} {28}},\ \bibinfo
  {pages} {816} (\bibinfo {year} {2010})}\BibitemShut {NoStop}%
  \BibitemOpen
\bibitem [{\citenamefont {Fan}\ \emph {et~al.}(2003)\citenamefont {Fan},
  \citenamefont {Suh},\ and\ \citenamefont {Joannopoulos}}]{Joannopoulos_2003}%
  \BibitemShut {NoStop}%
  \BibitemOpen
  \bibfield  {author} {\bibinfo {author} {\bibfnamefont {S.}~\bibnamefont
  {Fan}}, \bibinfo {author} {\bibfnamefont {W.}~\bibnamefont {Suh}}, \ and\
  \bibinfo {author} {\bibfnamefont {J.}~\bibnamefont {Joannopoulos}},\
  }\href@noop {} {\bibfield  {journal} {\bibinfo  {journal} {J. Opt. Soc. Am.
  A}\ }\textbf {\bibinfo {volume} {20}},\ \bibinfo {pages} {569} (\bibinfo
  {year} {2003})}\BibitemShut {NoStop}%
\bibitem [{\citenamefont {Howe}\ \emph {et~al.}(1981)\citenamefont {Howe},
  \citenamefont {Allan},\ and\ \citenamefont {Barnes}}]{howe1981}%
  \BibitemOpen
  \bibfield  {author} {\bibinfo {author} {\bibfnamefont {A.}~\bibnamefont
  {Howe}}, \bibinfo {author} {\bibfnamefont {D.~W.}\ \bibnamefont {Allan}}, \
  and\ \bibinfo {author} {\bibfnamefont {J.~A.}\ \bibnamefont {Barnes}},\
  }\href@noop {} {\bibfield  {journal} {\bibinfo  {journal} {Proc. 35th ann.
  symp. Freq. Control}\ } (\bibinfo {year} {1981})}\BibitemShut {NoStop}%
\bibitem [{\citenamefont {Maxin}(2014)}]{maxin2014}%
  \BibitemOpen
  \bibfield  {author} {\bibinfo {author} {\bibfnamefont {J.}~\bibnamefont
  {Maxin}},\ }\emph {\bibinfo {title} {Widely tunable optoelectronic oscillator
  and low noise for radar applications}},\ \href@noop {} {Ph.D. thesis},\
  \bibinfo  {school} {Universite Toulouse III Paul Sabatier} (\bibinfo {year}
  {2014})\BibitemShut {NoStop}%
\end{thebibliography}
\end{document}